\documentclass[12pt]{article}
\usepackage{graphicx} % Required for inserting images
\usepackage[margin=0.9in]{geometry}
\usepackage{amsthm,amsmath,amssymb,amscd,enumerate, graphicx,caption,subcaption,bm,booktabs,longtable,colortbl,array,makecell}
\usepackage{authblk}
\usepackage{placeins}
\DeclareMathOperator{\expit}{expit}
\usepackage{hyperref}
\usepackage{mdframed}
\usepackage{setspace}
\newcounter{algorithm}
\renewcommand{\thealgorithm}{\arabic{algorithm}}
\newenvironment{algorithm}[1][]{%
  \refstepcounter{algorithm}%
  \begin{mdframed}[linewidth=0.8pt,innertopmargin=6pt,innerbottommargin=6pt]%
  \noindent\textbf{Algorithm~\thealgorithm}%
  \ifx&#1&\else\textbf{: #1}\fi\par\smallskip\hrule\smallskip
}{%
  \end{mdframed}%
}
\newcommand{\AlgRequire}[1]{\textbf{Input:} #1\par}
\newcommand{\AlgEnsure}[1]{\textbf{Output:} #1\par\smallskip\hrule\smallskip}
\newcommand{\AlgState}[1]{\par\noindent\hspace{1em}#1}

\newcommand{\AlgReturn}[1]{\par\noindent\hspace{1em}\textbf{return} #1}
\usepackage{natbib}
\title{Robust variance estimation for multiple imputation using predictive mean matching or parametric imputation models}
\author[1,*]{Suyang Yu}
\author[2]{Lucy D'Agostino McGowan}
\author[3,4,5]{Brian~D. Williamson}
\affil[1]{Department of Statistics, University of Washington}
\affil[2]{Department of Statistical Sciences, Wake Forest University}
\affil[3]{Biostatistics Division, Kaiser Permanente Washington Health Research Institute}
\affil[4]{Department of Biostatistics, University of Washington}
\affil[5]{Vaccine and Infectious Disease Division, Fred Hutchinson Cancer Center}
\affil[*]{Corresponding Author. Email: syu3@ufl.edu. Current institution: University of Florida.}

\date{\today}

\begin{document}

\maketitle

\begin{abstract}
Multiple imputation (MI) is widely used for handling missing data in medical research, with variance commonly estimated using Rubin's rules (RR). RR is simple to apply, but under uncongeniality between the imputation and analysis procedures, its variance estimate can be either too small or too large, leading to miscalibrated confidence intervals. Robins and Wang (RW) developed an estimating-equation variance estimator that does not require congeniality under the conditions of their parametric imputation framework. Practical use of RW, however, has been limited by the lack of accessible software and by the additional variance calculations required for modern imputation workflows. We develop an R package, \texttt{rw}, construct RW variance components for normal and logistic conditional imputation models fitted through \texttt{mice}, and propose an approximate RW variance procedure for predictive mean matching (PMM). The PMM procedure preserves the original imputations and selected donors. For the variance calculation, we define differentiable donor probabilities and use expected analysis scores to approximate the cross-moment between analysis and imputation scores. Analysis scores are summed by recorded donor to account for donor reuse. We evaluate the proposed methods in benchmark simulations based on the original RW setting and in validation-study simulations motivated by electronic health record data. RR coverage was up to approximately 23 percentage points below the nominal 95\% level in some settings and above nominal in others. The proposed RW estimators generally produced coverage closer to the nominal level in larger samples under both parametric and PMM imputation, while retaining the same completed-data point estimates as RR.
\end{abstract}
% Keywords: missing data, multiple imputation, congeniality, uncongeniality, predictive mean matching

\section{Introduction}\label{introduction}

Multiple imputation (MI) is a standard tool in medical research for handling missing data in electronic health records, cohort studies, and clinical trials \citep{rubin1987multiple}. After creating several completed datasets, the standard approach is to average the completed-data point estimates and estimate their variance using Rubin's rules (RR), which combine the within-imputation variance with a scaled between-imputation variance \citep{rubin1987multiple, vanbuuren2011mice, sterne2009multiple, white2011multiple}. RR is convenient because it requires only the completed-data point estimates and their model-based standard errors, without access to the details of the imputation mechanism.

The usual justification for RR combines proper imputation, valid complete-data variance estimates, and congeniality between the imputation and analysis procedures \citep{meng1994multiple, wang1998large, xie2017dissecting}. Roughly, congeniality requires that the imputation procedure and complete-data analysis can be embedded in a common Bayesian model. Under uncongeniality, RR can either underestimate or overestimate the repeated-sampling variance \citep{bartlett2020bootstrap}. For example, it can be anti-conservative when the analysis includes interactions omitted from the imputation model and conservative when the imputation model uses information not included in the analysis. The direction depends on the particular imputation and analysis procedures, and the resulting confidence intervals may have coverage below or above the nominal level.

Two forms of uncongeniality are especially relevant in medical data analysis. The first is a structural mismatch between the imputation and analysis models, such as different assumptions about covariate effects, interactions, or error variances \citep{bartlett2015substantive, hughes2016comparison}. The second arises when study eligibility depends on imputed variables. In electronic health record (EHR) studies with error-prone measurements, the imputation model may be fitted using all subjects. After imputation, however, the analysis estimating equation may be evaluated only in a subset defined by the imputed value. \citet{giganti2020multiple}, hereafter GS, considered this setting in an HIV cohort study, where inclusion depended on antiretroviral therapy dispensation during the first month of care. Their simulations showed that RR could produce overly conservative intervals when subjects were excluded after imputation. Similar variance-estimation issues arise in related EHR validation-and-imputation analyses \citep{giganti2020dependent, shen2026missingness}.

\citet{robins2000inference} (hereafter RW) proposed an estimating-equation alternative to RR for variance estimation after multiple imputation. Instead of using the within- and between-imputation decomposition of RR, the RW estimator is built from three components: variability in the complete-data analysis score, variability propagated from estimation of the imputation-model parameters, and the cross-covariance between the two. Under the conditions of their parametric imputation framework, the estimator consistently estimates the repeated-sampling variance of the MI estimator without requiring the imputation and analysis models to be congenial or correctly specified. This result concerns the variance of the resulting estimator, not whether its probability limit equals the scientific target. Misspecification can still bias the point estimator, and changing the variance estimator does not remove that bias. We formalize this distinction in Section~\ref{motivating-setting}. Bootstrap procedures that repeat imputation and analysis within each resample provide another approach to inference under uncongeniality \citep{schomaker2018bootstrap, bartlett2020bootstrap}, but can be computationally expensive when the full imputation procedure must be repeated many times.

Applications of the RW estimator include simulation studies comparing MI variance estimators \citep{hughes2016comparison} and the EHR validation analyses of GS and colleagues \citep{giganti2020dependent, giganti2020multiple}. GS found that for the estimated association between baseline CD4 count and subsequent poor outcomes, the standard error based on the RW method was approximately $29\%$ smaller than the RR estimate.  These implementations are useful in specific settings, but broader use requires a practical way to obtain the variance components from fitted imputation and analysis models. Two commonly used imputation procedures create additional complications. First, multiple imputation with chained equations (MICE) fits a sequence of conditional models that need not correspond to a common joint distribution \citep{vanbuuren2011mice}. The scores from these conditional models must therefore be distinguished from the score of a joint imputation model. Second, predictive mean matching (PMM), a commonly-used imputation method and the default in the \texttt{mice} R package \citep{vanbuuren2011mice},  imputes a missing value by copying the value of a matched observed donor rather than drawing from a parametric conditional distribution \citep{schenker1995partially, andridge2010review}. Its discrete donor-selection rule and the possible reuse of the same donor by multiple recipients require additional adjustments to the variance calculation.

We develop an R package, \texttt{rw}, for RW variance estimation in \texttt{mice}-based workflows \citep{rwpkg}. For normal and logistic conditional imputation models, we construct the required RW components by combining the score and influence contributions from the conditional fits. We also propose an approximate RW variance procedure for PMM that preserves the imputed values and selected donors produced by \texttt{mice}. For this variance calculation, we define differentiable donor probabilities and use expected analysis scores to approximate the cross-moment between analysis and imputation scores. We sum analysis scores by recorded donor to account for dependence induced by donor reuse. Related approaches to variance estimation for hot-deck and nearest-neighbor imputation have been studied in survey sampling \citep{rao1992jackknife,chen2002common,kim2004fractional,yang2020pmm}. We evaluate the proposed estimators in simulations by comparing estimated standard errors and confidence-interval coverage with those from RR, while keeping the completed-data point estimates unchanged.

The remainder of the paper is organized as follows. Section~\ref{variance-estimation} introduces the data structure, notation, MI point estimator, and RW variance estimator. Section~\ref{improved-variance} develops the proposed MICE and PMM procedures, and Section~\ref{software-implementation} describes their implementation. Section~\ref{simulation-studies} compares the proposed estimators with RR in the original RW simulation setting and the GS measurement-error validation setting. Section~\ref{discussion} discusses limitations and directions for future research.

\section{Variance estimation with imputed data}\label{variance-estimation}
% 2.1 Data structure and notation
\subsection{Data structure and notation}\label{data-notation}
We observe independent and identically distributed data \(O_1,\ldots,O_n\), where each \(O_i\) records the observed components for subject \(i\) together with its missingness pattern. We write \(Z_i\) for the complete data, including both observed and missing components. Throughout, \(P_n\) denotes the empirical average, \(P_n f = n^{-1}\sum_{i=1}^n f_i\). Let \(\beta_0\in\mathbb{R}^q\) denote the complete-data population parameter, defined by \(E\{U(Z;\beta_0)\}=0\), where \(U(\cdot\,;\beta)\) is the complete-data estimating function and uniqueness is assumed. This definition also applies when the analysis model is misspecified. When \(Z\) is partially missing, the estimating function cannot be evaluated directly, and multiple imputation replaces the missing components by draws from an imputation model.

In this section we consider parametric imputation, the setting in which the RW estimator was developed. The imputation model has parameter \(\psi\in\mathbb{R}^c\), estimated from the observed data by \(\hat\psi\) solving \(\sum_{i=1}^n S_{\mathrm{obs},i}(\psi)=0\), with probability limit \(\psi^\ast\). The influence function of \(\hat\psi\) is the mean-zero function \(d_i\) satisfying \(\hat\psi-\psi^\ast = n^{-1}\sum_{i=1}^n d_i + o_p(n^{-1/2})\). Drawing from the fitted imputation model \(m\) times produces \(m\) completed datasets, indexed by \(p=1,\ldots,m\). We write \(Z_i^{(p)}\) for subject \(i\)'s completed record in imputation \(p\) and \(U_i^{(p)}(\beta)=U\{Z_i^{(p)};\beta\}\) for its estimating-function contribution. PMM instead fills in missing values using copies from an observed donor with a similar predicted mean \citep{schenker1995partially, andridge2010review}; Section~\ref{extension-2} introduces variance calculation and donor notation for PMM.
%======================================================================
% 2.2 Point estimation with imputed data
%======================================================================
\subsection{Point estimation using multiple imputation}\label{point-estimation}\label{motivating-setting}
The point estimate following multiple imputation is obtained by fitting the analysis model to each of the \(m\) completed datasets and averaging the resulting estimates. We define $\hat{\beta}^{(p)}$ as the resulting point estimate from solving $P_nU^{(p)}(\hat\beta^{(p)})=0$ imputed dataset $p$. Then the final point estimate is $\hat\beta=m^{-1}\sum_{p=1}^m\hat\beta^{(p)}.$

The MI estimator may converge to a limit different from the complete-data target \(\beta_0\). Denote this probability limit, when it exists, by \(\beta^\ast\). The value of \(\beta^\ast\) depends on the imputation and analysis procedures and on the missingness mechanism. Under regularity conditions, \(\beta^\ast=\beta_0\) if the completed-data estimating function has expectation zero at \(\beta_0\) under the limiting imputation procedure. Imputation-model misspecification does not necessarily imply \(\beta^\ast\ne\beta_0\). For example, when only the response is imputed and the predictors are fully observed, misspecifying the imputation error variance can leave the limiting linear regression coefficient unchanged if the conditional mean is correctly specified.

Coverage of \(\beta^\ast\) concerns the estimator's limiting target, whereas coverage of \(\beta_0\) also reflects any difference between the two targets. In the simulations, we approximate coverage of \(\beta^\ast\) by using the mean point estimate across replications in place of the limiting target.
%======================================================================
% 2.3 Variance estimation using Rubin's rules
%======================================================================
\subsection{Variance estimation using Rubin's rules}\label{rubins-rules}
The standard approach estimates the variance of \(\hat\beta\) by Rubin's rules (RR). Let \(\widehat W^{(p)}\) be the corresponding complete-data (model-based) variance estimate of \(\hat\beta^{(p)}\). RR combines the average within-imputation variance,
\[
\bar W = \frac{1}{m}\sum_{p=1}^m \widehat W^{(p)},
\]
with the between-imputation variance,
\[
\widehat B
=
\frac{1}{m-1}\sum_{p=1}^m
\bigl(\hat\beta^{(p)}-\hat\beta\bigr)
\bigl(\hat\beta^{(p)}-\hat\beta\bigr)^\top,
\]
to give
\[
\widehat V_{\mathrm{RR}}
=
\bar W + \Bigl(1+\frac{1}{m}\Bigr)\widehat B .
\]
Here, \(\bar W\) averages the complete-data variance estimates across the \(m\) datasets, and \(\widehat B\) measures variation in the point estimates across imputations. The factor \(1+1/m\) accounts for the additional variation associated with finite \(m\). The usual justification for RR combines proper imputation, valid complete-data variance estimates, and congenial imputation and analysis procedures \citep{rubin1987multiple, meng1994multiple, wang1998large}. Under uncongeniality, RR can overestimate or underestimate the repeated-sampling variance.

\begin{figure}[ht]
  \centering
  \includegraphics[width=\textwidth]{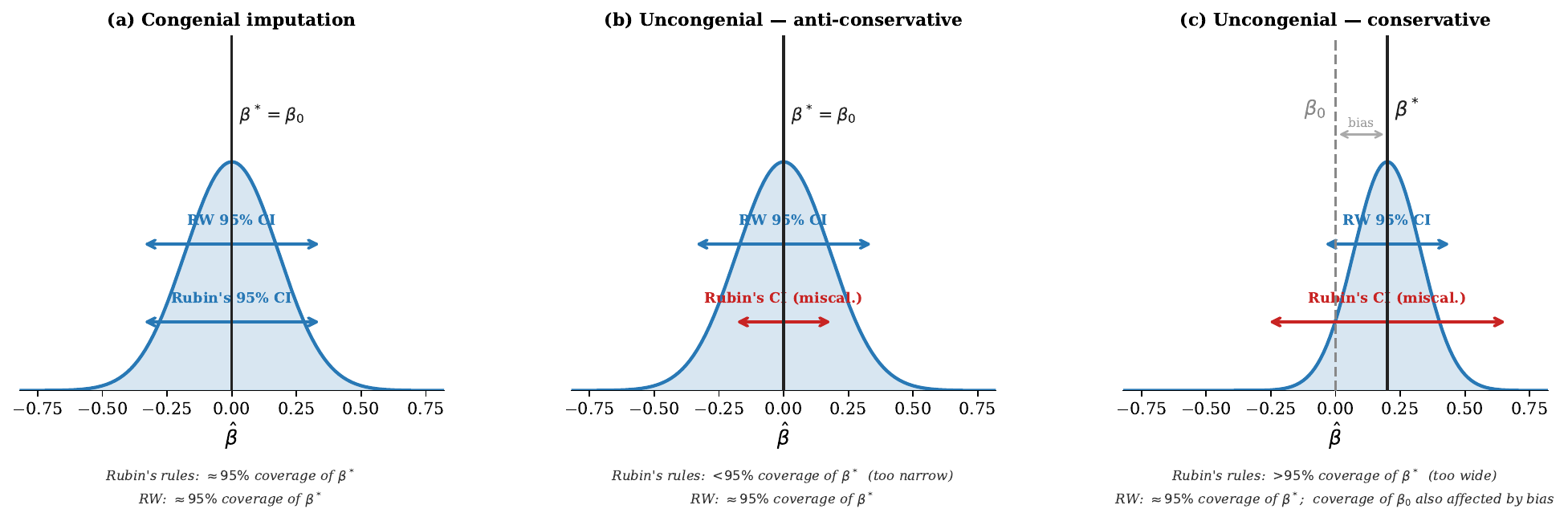}
  \caption{Conceptual illustration of confidence-interval coverage for multiply imputed estimators. Each panel shows the sampling distribution of \(\hat\beta\) under repeated study replications, the probability limit \(\beta^\ast\) (solid vertical line), and, in panel~(c), the complete-data population parameter \(\beta_0\) (dashed vertical line). Horizontal brackets represent 95\% confidence intervals based on RR and RW variance estimates. In panel~(a), both attain approximately 95\% coverage of \(\beta^\ast\). In panel~(b), the RR interval is too narrow and undercovers \(\beta^\ast\). In panel~(c), the RR interval is too wide and overcovers \(\beta^\ast\); when \(\beta^\ast\ne\beta_0\), coverage of \(\beta_0\) also depends on the difference between the two targets.}
  \label{fig:schematic}
\end{figure}

Figure~\ref{fig:schematic} illustrates how variance misspecification affects confidence-interval coverage. It also distinguishes coverage of the limiting target \(\beta^\ast\) from coverage of the complete-data target \(\beta_0\). The RW intervals are shown schematically under conditions for which the RW variance formula is valid.
%======================================================================
% 2.4 Variance estimation using Robins-Wang
%======================================================================
\subsection{Variance estimation using Robins--Wang}
\label{robinswang-variance-components}
The Robins--Wang (RW) estimator accounts for variation in the analysis scores, uncertainty in the fitted imputation model, and their cross-covariance. Under its regularity conditions, the parametric RW estimator remains valid without congeniality \citep{robins2000inference}.

In the original RW construction, each subject's missing components are drawn from \(f(Z_{\mathrm{mis}}\mid O;\hat\psi)\), with \(\hat\psi\) fixed across imputations. Conditional on the observed data, these draws are independent across subjects and imputations. The analysis estimate, denoted by \(\tilde\beta\), solves
\[
P_n\bar U(\tilde\beta)=0,\qquad
\bar U_i(\beta)=\frac{1}{m}\sum_{p=1}^m U_i^{(p)}(\beta).
\]
Section~\ref{improved-variance} gives the components used with chained and PMM imputations and the pooled estimate \(\hat\beta\) defined in Section~\ref{point-estimation}.

Besides the analysis score, RW requires the missing-value score and an estimated influence contribution for \(\hat\psi\):
\[
S_{\mathrm{mis},i}^{(p)}
=\left.\frac{\partial}{\partial\psi}
\log f(Z_{\mathrm{mis},i}^{(p)}\mid O_i;\psi)
\right|_{\psi=\hat\psi},
\]
\[
d_i
=
-\left\{P_n\left.
\frac{\partial S_{\mathrm{obs},i}(\psi)}{\partial\psi^\top}
\right|_{\psi=\hat\psi}\right\}^{-1}
S_{\mathrm{obs},i}(\hat\psi).
\]
For fully observed records, \(S_{\mathrm{mis},i}^{(p)}=0\). Records that do not contribute to the observed-data score have \(d_i=0\). The contribution \(d_i\) is the same in every imputation. Below, we write \(U_i^{(p)}=U_i^{(p)}(\tilde\beta)\) and \(\bar U_i=m^{-1}\sum_p U_i^{(p)}\). Following the notation of \citet{giganti2020multiple}, the sandwich variance estimator is
\[
\widehat\Gamma
=
\frac{1}{n}\widehat\tau^{-1}\widehat\Delta\,\bigl(\widehat\tau^{-1}\bigr)^{\!\top},
\qquad
\widehat\Delta
=
\widehat\Omega
+\widehat\kappa\widehat\Lambda\widehat\kappa^\top
+\widehat C .
\]
Here \(\widehat\Delta\) estimates the covariance of the first-order score contributions, and \(\widehat\tau\) is the negative average derivative of the completed-data estimating equation with respect to \(\beta\),
\[
\widehat\tau
=
-\frac{1}{nm}\sum_{p=1}^m\sum_{i=1}^n
\left.\frac{\partial U_i^{(p)}(\beta)}{\partial\beta^\top}
\right|_{\beta=\tilde\beta}.
\]
We now describe the three sources of variance in $\hat{\Delta}$.

\medskip
\noindent\textit{Source 1: analysis-score covariance.}\enskip The first piece is the outer product of each subject's imputation-averaged analysis score \(\bar U_i\), retaining both between-subject variation and residual imputation noise for finite \(m\):
\[
\widehat\Omega
=
\frac{1}{n}\sum_{i=1}^n
\bar U_i\bar U_i^\top.
\]

\medskip
\noindent\textit{Source 2: imputation-parameter uncertainty.}\enskip The matrix \(\widehat\kappa\) estimates the sensitivity of the expected analysis score to \(\psi\), while \(\widehat\Lambda\) estimates the covariance of the influence contributions \(d_i\):
\[
\widehat\kappa
=
\frac{1}{nm}\sum_{p=1}^m\sum_{i=1}^n
U_i^{(p)}S_{\mathrm{mis},i}^{(p)\top},
\qquad
\widehat\Lambda
=
\frac{1}{n}\sum_{i=1}^n d_i d_i^\top.
\]
Their contribution to \(\widehat\Delta\) is \(\widehat\kappa\widehat\Lambda\widehat\kappa^\top\).

\medskip
\noindent\textit{Source 3: cross-covariance.}\enskip The same observed records contribute to both the analysis scores and the fitted imputation model. Their dependence is accounted for by
\[
\widehat C
=
\frac{1}{n}
\Bigl[
\widehat\kappa
\Bigl\{\sum_{i=1}^n d_i\,\bar U_i^\top\Bigr\}
+
\Bigl\{\sum_{i=1}^n \bar U_i\,d_i^\top\Bigr\}
\widehat\kappa^\top
\Bigr].
\]
This cross-covariance term can increase or decrease the total variance.

\section{Improved RW variance estimation with multiply-imputed data}
\label{improved-variance}
This section applies the three-term RW variance formula in Section~\ref{robinswang-variance-components} to chained conditional imputation and PMM. In both cases, the point estimate remains the mean of the completed-data estimates. Section~\ref{extension-1} calculates the score and influence contributions for normal and logistic conditional models and combines them into the vectors \(S_{\mathrm{mis}}\) and \(d\). Section~\ref{extension-2} gives the PMM calculation: donor probabilities and expected analysis scores \(G_{ij}\) determine the PMM columns of \(\widehat\kappa\), and scores summed by recorded donor determine \(\widehat\Omega\) and \(\widehat C\). We denote the resulting estimators by RW-MICE and RW-PMM, respectively.
\subsection{RW variance for chained conditional imputation models}
\label{extension-1}
In the MICE specifications considered here, a normal or logistic conditional model is specified for each variable \(r=1,\ldots,R\) requiring imputation.

For each completed dataset \(p\), running \texttt{mice} with \texttt{tasks = "train"} records the parameter draw \(\psi_r^{(p)}\) used to impute variable \(r\) in the final MICE iteration. For a normal model, \(\psi_r=(\eta_r^\top,\sigma_r^2)^\top\), where \(\eta_r\) contains the regression coefficients and \(\sigma_r^2\) is the residual variance; for a logistic model, \(\psi_r\) consists of the regression coefficients.

\subsubsection{Scores and parameter estimation}
Let \(Y_{r,i}^{(p)}\) denote the observed or imputed value of variable \(r\) for subject \(i\) in completed dataset \(p\), and let \(x_{r,i}^{(p)}\) contain the predictors used in its conditional model. If a predictor is updated later in the final iteration, its completed value may differ from the value used to impute variable \(r\). The score calculation uses the completed-data values together with the recorded parameter draw \(\psi_r^{(p)}\).

Write the conditional model as \(f_r\{Y_{r,i}^{(p)}\mid x_{r,i}^{(p)};\psi_r\}\), with score
\[
s_{r,i}^{(p)}(\psi_r)
=
\frac{\partial}{\partial\psi_r}
\log f_r\{Y_{r,i}^{(p)}\mid x_{r,i}^{(p)};\psi_r\}.
\]
Let \(\delta_{r,i}=1\) if variable \(r\) is observed for subject \(i\), and \(0\) otherwise. The score contributions from observed and imputed responses are
\[
S_{r,\mathrm{obs},i}^{(p)}
= \delta_{r,i}\,s_{r,i}^{(p)}(\psi_r^{(p)}),
\qquad
S_{r,\mathrm{mis},i}^{(p)}
= (1-\delta_{r,i})\,s_{r,i}^{(p)}(\psi_r^{(p)}),
\]
respectively. The derivative matrix corresponding to the observed-response scores is
\[
H_r^{(p)}
=
\frac{1}{n}\sum_{i=1}^n
\delta_{r,i}
\left.
\frac{\partial s_{r,i}^{(p)}(\psi_r)}
{\partial\psi_r^\top}
\right|_{\psi_r=\psi_r^{(p)}},
\]
where derivatives are taken with the completed data held fixed. For invertible \(H_r^{(p)}\), the estimated contribution of subject \(i\) to fitting conditional model \(r\) is
\[
d_{r,i}^{(p)}
=
-\{H_r^{(p)}\}^{-1}S_{r,\mathrm{obs},i}^{(p)}.
\]
Arrange the contributions in the same parameter order for all imputations:
\[
S_{\mathrm{mis},i}^{(p)}
=
\bigl(
S_{1,\mathrm{mis},i}^{(p)\top},\ldots,S_{R,\mathrm{mis},i}^{(p)\top}
\bigr)^\top,
\qquad
d_i^{(p)}
=
\bigl(
d_{1,i}^{(p)\top},\ldots,d_{R,i}^{(p)\top}
\bigr)^\top.
\]

Let \(\psi_r^\ast\) denote the probability limit of the parameter estimate for conditional model \(r\), and assume that each recorded draw satisfies
\[
\psi_r^{(p)}-\psi_r^\ast=O_p(n^{-1/2}).
\]
Under the regularity conditions in Appendix~\ref{appendix mice}, substituting the recorded draws for their probability limits changes the empirical averages below by \(o_p(1)\).

\subsubsection{RW components for the MICE estimator}
Let \(\hat\beta^{(p)}\) solve the analysis estimating equation in completed dataset \(p\). As above, the multiple-imputation point estimate is the average across imputations:
\[
P_nU^{(p)}(\hat\beta^{(p)})=0,\qquad
\hat\beta=\frac{1}{m}\sum_{p=1}^m\hat\beta^{(p)}.
\]
Write the fitted analysis scores and the averages across imputations as
\[
U_i^{(p)}=U_i^{(p)}(\hat\beta^{(p)}),\qquad
\bar U_i=\frac{1}{m}\sum_{p=1}^mU_i^{(p)},\qquad
\text{ and } \bar d_i=\frac{1}{m}\sum_{p=1}^m d_i^{(p)}.
\]
The negative average derivative of the analysis score is
\[
\widehat\tau=-\frac{1}{nm}\sum_{p=1}^m\sum_{i=1}^n
\left.\frac{\partial U_i^{(p)}(\beta)}{\partial\beta^\top}
\right|_{\beta=\hat\beta^{(p)}}.
\]
Any estimated residual variance in the analysis model is held fixed when differentiating. For analyses restricted to a subset, excluded subjects contribute zero scores and derivatives within each completed dataset. Averages over subjects use the full sample size \(n\) as the denominator.

The products in \(\widehat\kappa\) and \(\widehat\Lambda\) are formed within each imputation, using its recorded parameter draws, and then averaged. In \(\widehat\Omega\) and \(\widehat C\), each subject's contributions are averaged across imputations before forming the products:
\begin{align*}
\widehat\Omega
&=\frac{1}{n}\sum_{i=1}^n \bar U_i\bar U_i^\top,
\\[4pt]
\widehat\kappa
&=\frac{1}{nm}\sum_{p=1}^m\sum_{i=1}^n
U_i^{(p)}S_{\mathrm{mis},i}^{(p)\top},
\qquad
\widehat\Lambda
=\frac{1}{nm}\sum_{p=1}^m\sum_{i=1}^n
d_i^{(p)}d_i^{(p)\top},
\\[4pt]
\widehat C
&=\frac{1}{n}\sum_{i=1}^n
\left\{
\widehat\kappa\bar d_i\bar U_i^\top
+\bar U_i\bar d_i^\top\widehat\kappa^\top
\right\}.
\end{align*}
Subjects with observed responses in more than one conditional model contribute to several parameter estimates. Their cross-products enter the corresponding off-diagonal submatrices of \(\widehat\Lambda\). The resulting variance estimator for \(\hat\beta\) is
\[
\widehat\Gamma_{\mathrm{MICE}}
=\frac{1}{n}\widehat\tau^{-1}
\bigl(\widehat\Omega+\widehat\kappa\widehat\Lambda\widehat\kappa^\top
+\widehat C\bigr)(\widehat\tau^{-1})^\top.
\]

In the GS setting, both conditional models are fitted to validation records with observed predictors. Appendix~\ref{appendix mice} uses this structure to derive the large-\(m\) variance of the pooled estimator.

\subsection{RW variance estimation with predictive mean matching}
\label{extension-2}
PMM selects a donor uniformly from the \(k\) observed records whose predicted means are closest to the recipient's predicted mean and copies that donor's value. Its selection probabilities are piecewise constant in the matching coefficients and change discontinuously when the donor pool changes. This presents a challenge for the RW variance, because there is not a clear set of imputation model parameters that index the imputation distribution. We propose to approximate the RW cross-moment \(\kappa=E\{US_{\mathrm{mis}}^\top\}\) using differentiable donor probabilities and expected analysis scores.

Dependence also arises when several records use the same donor. To account for the shared donor value, we sum their analysis scores before calculating \(\widehat\Omega\) and \(\widehat C\). The matching regression supplies the contributions \(d_i\) from parameter estimation.

\subsubsection{Donor probabilities for variance estimation}
\label{smooth-copied-donor}
Consider one continuous variable \(A\) imputed by PMM, with observed and missing sets \(\mathcal O_A\) and \(\mathcal M_A\). Let \(x_i\) contain the fully observed matching predictors, including an intercept. Under the default matching rule in \texttt{mice}, observed donors use the fitted regression coefficient \(\hat\theta^{(p)}\), whereas recipients use a coefficient draw \(\theta^{(p)}\). Their matching predictions are
\[
a_j^{(p)}=x_j^\top\hat\theta^{(p)},\qquad
t_i^{(p)}=x_i^\top\theta^{(p)}.
\]
Let \(\mathcal N_i^{(p)}\subset\mathcal O_A\) contain the indices of the \(k\) donors whose predictions are closest to \(t_i^{(p)}\). With selected donor index \(J_i^{(p)}\), imputation follows
\[
P\{J_i^{(p)}=j\mid\text{fitted model and coefficient draw}\}
=\frac{\mathbf 1\{j\in\mathcal N_i^{(p)}\}}{k},
\qquad A_i^{(p)}=A_{J_i^{(p)}}^{\mathrm{obs}}.
\]
This rule generates the completed datasets used for analysis. The following probability calculation is used to approximate the variance.

Suppress the superscript \(p\) temporarily and order the \(N=|\mathcal O_A|\) donor predictions as \(a_{(1)}<\cdots<a_{(N)}\). The derivatives below assume distinct predictions and a locally unchanged ordering. Donor \((r)\) is among the \(k\) nearest donors to a scalar prediction \(t\) when \(\ell_{(r)}<t<u_{(r)}\), apart from ties at the endpoints, where
\[
\ell_{(r)}=
\begin{cases}
-\infty,&r\leq k,\\
(a_{(r-k)}+a_{(r)})/2,&r>k,
\end{cases}
\qquad
u_{(r)}=
\begin{cases}
(a_{(r)}+a_{(r+k)})/2,&r+k\leq N,\\
\infty,&r+k>N.
\end{cases}
\]
These midpoints mark where a competing donor enters or leaves the nearest \(k\) positions.

Define \(T_i\sim N(t_i,s_i^2)\), where
\[
s_i^2=x_i^\top\Sigma x_i,\qquad \Sigma=\sigma^2 V.
\]
Here \(\sigma\) is the residual standard-deviation draw from the PMM regression, and \(V\) is its fitted coefficient covariance factor. For a full-rank unregularized fit, \(V=(X_{\mathcal O_A}^\top X_{\mathcal O_A})^{-1}\); the implementation uses the factor from the \texttt{mice} regression fit. The variance \(s_i^2\) describes variation in the recipient's predicted mean induced by \(\Sigma\). Averaging the uniform selection rule over \(T_i\) gives
\begin{equation}
p_{ij}=\frac{1}{k}
\left\{\Phi\!\left(\frac{u_j-t_i}{s_i}\right)
-\Phi\!\left(\frac{\ell_j-t_i}{s_i}\right)\right\}.
\label{pmm-probability}
\end{equation}
The term in braces is \(P(\ell_j<T_i<u_j)\), the probability that donor \(j\) belongs to the pool under this normal distribution. Division by \(k\) accounts for uniform selection within the pool, giving \(\sum_{j\in\mathcal O_A}p_{ij}=1\).

To differentiate, shift both \(\hat\theta\) and \(\theta\) by the same vector \(\delta\), with \(\Sigma\) held fixed. The derivative of the recipient prediction is \(x_i\), and the derivatives of finite endpoints are
\[
\dot\ell_{(r)}=(x_{(r-k)}+x_{(r)})/2,\qquad
\dot u_{(r)}=(x_{(r)}+x_{(r+k)})/2.
\]
Writing \(z_{ij}^{u}=(u_j-t_i)/s_i\) and \(z_{ij}^{\ell}=(\ell_j-t_i)/s_i\), differentiation gives
\begin{equation}
\dot p_{ij}:=\left.\nabla_\delta p_{ij}(\delta)\right|_{\delta=0}
=\frac{\phi(z_{ij}^{u})(\dot u_j-x_i)
-\phi(z_{ij}^{\ell})(\dot\ell_j-x_i)}{k s_i}.
\label{pmm-probability-derivative}
\end{equation}
Terms corresponding to infinite endpoints are zero, and the derivatives sum to zero over donors. An intercept shift moves all predictions equally, so the intercept coordinate of \(\dot p_{ij}\) is zero.

\subsubsection{Expected analysis scores and the PMM cross-moment}
For recipient \(i\) and candidate donor \(j\), calculate the expected analysis score by averaging over \(A_j^\dagger\sim N(a_j^{(p)},(\sigma^{(p)})^2)\). If a subsequent variable \(Y_i^\dagger\) is imputed using \(A\) as a predictor, average also over its recorded conditional distribution evaluated at \(A_j^\dagger\). The recipient's observed covariates and outcomes are held fixed, and the analysis score includes its original subset indicator. Define
\[
G_{ij}^{(p)}
=E\!\left[
U_i\{A_j^\dagger,Y_i^\dagger;\hat\beta^{(p)}\}
\mid\text{fitted models and observed data}
\right].
\]
The two normal averages enter different parts of the calculation: \(s_i^2\) determines the averaging over recipient predictions in \(p_{ij}\), while \((\sigma^{(p)})^2\) determines the averaging over donor values in \(G_{ij}\).

For the probabilities in \eqref{pmm-probability}, define \(S_{ij}^{(p)}=\dot p_{ij}^{(p)}/p_{ij}^{(p)}\) when \(p_{ij}^{(p)}>0\). Then
\[
\sum_{j\in\mathcal O_A}p_{ij}^{(p)}G_{ij}^{(p)}S_{ij}^{(p)\top}
=\sum_{j\in\mathcal O_A}G_{ij}^{(p)}\dot p_{ij}^{(p)\top}.
\]
The PMM columns of \(\widehat\kappa\) are approximated by
\begin{equation}
\widehat\kappa_{\mathrm{PMM}}
=\frac{1}{nm}\sum_{p=1}^m\sum_{i\in\mathcal M_A}
\sum_{j\in\mathcal O_A}G_{ij}^{(p)}\dot p_{ij}^{(p)\top}.
\label{pmm-kappa}
\end{equation}
The derivative acts on \(p_{ij}\), with the models used in \(G_{ij}\) held fixed. Appendix~\ref{appendix pmm} expresses each recipient's contribution as a weighted sum of expected-score differences between neighboring donors and derives its limit when the matching coefficients are known and donor predictions become dense.

For example, if \(A\) is the response in a linear analysis with fully observed design vector \(z_i\) and fitted residual variance \(\hat v^{(p)}\), then
\[
U_i^{(p)}=\frac{z_i\{A_i^{(p)}-z_i^\top\hat\beta^{(p)}\}}{\hat v^{(p)}},
\qquad
G_{ij}^{(p)}=\frac{z_i\{a_j^{(p)}-z_i^\top\hat\beta^{(p)}\}}{\hat v^{(p)}}.
\]
For a logistic analysis of \(D\) on \(A\) restricted to \(A>c_0\), let \(z(a)=(1,a)^\top\) and \(\pi_\beta(a)=\expit\{z(a)^\top\beta\}\). Let \(\rho_i^{(p)}(a)\) be the probability of \(D_i=1\) under its recorded logistic imputation model, evaluated at \(A=a\); if \(D_i\) is observed, set \(\rho_i^{(p)}(a)=D_i\). The required expectation is
\begin{equation}
G_{ij}^{(p)}=\int_{c_0}^{\infty}
z(a)\{\rho_i^{(p)}(a)-\pi_{\hat\beta^{(p)}}(a)\}
\frac{1}{\sigma^{(p)}}
\phi\!\left(\frac{a-a_j^{(p)}}{\sigma^{(p)}}\right)\,da.
\label{pmm-binomial-score}
\end{equation}
The lower integration limit incorporates the analysis restriction \(A>c_0\).

\subsubsection{Accounting for donor reuse}
\label{donor-source-analysis-scores}
Across repeated samples, variation in an observed donor value affects the analysis scores of its recipients. For each \(j\in\mathcal O_A\), sum the donor's score and the scores of recipients who use that donor:
\[
\widetilde U_j^{(p)}(\beta)
=
U_j^{(p)}(\beta)
+
\sum_{\substack{i\in\mathcal M_A\\J_i^{(p)}=j}} U_i^{(p)}(\beta).
\]
Set \(\widetilde U_i^{(p)}(\beta)=0\) for \(i\in\mathcal M_A\). Each \(U_i^{(p)}(\beta)\) uses subject \(i\)'s own covariates and eligibility indicator. Summing over all records gives
\[
\sum_{i=1}^n U_i^{(p)}(\beta)
=
\sum_{i=1}^n \widetilde U_i^{(p)}(\beta).
\]
This equality holds for every \(\beta\) and also after differentiation with the donor assignments held fixed, giving the same point estimate and analysis-score derivative. If \(\mathcal I_j^{(p)}\) contains donor \(j\) and its recipients, the outer product is
\[
\widetilde U_j^{(p)}\widetilde U_j^{(p)\top}
=\sum_{i\in\mathcal I_j^{(p)}}U_i^{(p)}U_i^{(p)\top}
+\sum_{\substack{i,\ell\in\mathcal I_j^{(p)}\\i\ne\ell}}
U_i^{(p)}U_\ell^{(p)\top}.
\]
The second sum contains the donor--recipient and recipient--recipient cross-products. Shared donor values also motivate variance estimators for hot-deck and nearest-neighbor imputation \citep{rao1992jackknife,chen2002common,kim2004fractional,yang2020pmm}.

For a linear analysis with a correctly specified conditional mean, recipients that match closely share the donor's residual, weighted by their own analysis covariates. In the logistic example above, the shared value of \(A\) enters each recipient's mean score, while the subsequent binary draws contribute additional variation. Appendix~\ref{appendix pmm} derives these expressions and their conditional covariance.

\subsubsection{RW components for use with PMM}
For imputation \(p\), evaluate the analysis scores \(U_i^{(p)}\) at \(\hat\beta^{(p)}\) and sum them by the recorded donor to obtain \(\widetilde U_i^{(p)}\). Form these sums separately within each imputation, then average:
\[
\bar{\widetilde U}_i
=
\frac{1}{m}\sum_{p=1}^m \widetilde U_i^{(p)},
\qquad
\widehat\Omega_{\mathrm{PMM}}
=
\frac{1}{n}\sum_{i=1}^n
\bar{\widetilde U}_i\bar{\widetilde U}_i^\top.
\]
The PMM columns of \(\widehat\kappa\) are given by \eqref{pmm-kappa}. For additional normal or logistic imputation models, let \(S_{\mathrm{par},i}^{(p)}\) contain their missing-value scores. The corresponding columns are calculated using the completed-data analysis scores:
\[
\widehat\kappa_{\mathrm{par}}
=
\frac{1}{nm}\sum_{p=1}^m\sum_{i=1}^n
U_i^{(p)}S_{\mathrm{par},i}^{(p)\top}.
\]
For the PMM matching regression, linearizing the score equation gives
\[
H_{\mathrm{PMM}}^{(p)}=-\frac{1}{n(\sigma^{(p)})^2}
\sum_{j\in\mathcal O_A}x_jx_j^\top,\qquad
d_{\mathrm{PMM},i}^{(p)}=
\begin{cases}
-\{H_{\mathrm{PMM}}^{(p)}\}^{-1}\dfrac{x_i(A_i-a_i^{(p)})}{(\sigma^{(p)})^2},
&i\in\mathcal O_A,\\[4pt]
0,&i\in\mathcal M_A.
\end{cases}
\]
This contribution contains one coordinate for each matching regression coefficient. Combine it with the normal or logistic contributions from Section~\ref{extension-1} to form \(d_i^{(p)}\), in the same parameter order as the columns of \(\widehat\kappa\). Define
\[
\widehat\Lambda=\frac{1}{nm}\sum_{p=1}^m\sum_{i=1}^n
d_i^{(p)}d_i^{(p)\top},\qquad
\bar d_i=\frac{1}{m}\sum_{p=1}^m d_i^{(p)}.
\]
Using the same donor sums as in \(\widehat\Omega_{\mathrm{PMM}}\), the cross-covariance term is
\[
\widehat C_{\mathrm{PMM}}
=
\frac{1}{n}\sum_{i=1}^n
\left\{
\widehat\kappa\bar d_i\bar{\widetilde U}_i^\top
+\bar{\widetilde U}_i\bar d_i^\top\widehat\kappa^\top
\right\}.
\]
Combining the components gives
\[
\widehat\Delta_{\mathrm{PMM}}
=
\widehat\Omega_{\mathrm{PMM}}
+
\widehat\kappa\,\widehat\Lambda\,\widehat\kappa^\top
+
\widehat C_{\mathrm{PMM}},
\qquad
\widehat\Gamma_{\mathrm{PMM}}
=
\frac{1}{n}
\widehat\tau^{-1}
\widehat\Delta_{\mathrm{PMM}}
\bigl(\widehat\tau^{-1}\bigr)^{\!\top},
\]
where
\[
\widehat\tau=-\frac{1}{nm}\sum_{p=1}^m\sum_{i=1}^n
\left.\frac{\partial U_i^{(p)}(\beta)}{\partial\beta^\top}
\right|_{\beta=\hat\beta^{(p)}}.
\]
The estimator uses the original completed-data analyses and their pooled point estimate \(\hat\beta=m^{-1}\sum_p\hat\beta^{(p)}\). Appendix~\ref{appendix pmm} gives the variance decomposition for these components and supporting calculations under the stated linear and logistic models.

\section{Software Implementation}\label{software-implementation}
The R package \texttt{rw} implements the variance estimators in Sections~\ref{extension-1} and~\ref{extension-2} for linear and logistic regression analyses. Imputations are generated with \texttt{mice} \citep{vanbuuren2011mice}, using \texttt{tasks = "train"} to save the fitted models and parameter draws. Normal and logistic imputations use \texttt{norm} and \texttt{logreg}, respectively. For a continuous variable imputed by PMM, \texttt{pmmrw} uses the donor draws from \texttt{mice} and also records the selected donor IDs and matching-regression information.

The package follows a similar structure to \texttt{mice}. The function \texttt{with\_rw()} fits the specified analysis to each completed dataset and evaluates the analysis scores and parameter-estimation contributions. The function \texttt{pool\_rw()} averages the coefficient estimates and calculates their RW covariance matrix. Algorithm~\ref{alg:rw-mice-pmm} gives the calculation for parametric imputation and for imputation involving one continuous PMM variable. Code and documentation are available at \url{https://github.com/LucyMcGowan/rw}.

\begin{algorithm}[RW variance estimation with parametric or PMM imputation]
\label{alg:rw-mice-pmm}
\AlgRequire{Observed data $\{O_i\}_{i=1}^n$; imputation models for variables $r=1,\ldots,R$; number of imputations $m$; analysis estimating function $U$; donor pool size $k$ if a continuous variable $A$ uses PMM.}
\AlgEnsure{Pooled estimate $\hat\beta$ and RW covariance estimate $\widehat\Gamma$.}
\AlgState{\textbf{Step 1 (Impute).} Generate $m$ completed datasets with \texttt{mice}. Record the model and parameter draw from the last update of each variable in the final iteration. For PMM, also record the selected donor IDs $J_i^{(p)}$, matching coefficients $\hat\theta^{(p)}$ and draws $\theta^{(p)}$, and residual scale $\sigma^{(p)}$.}
\AlgState{\textbf{Step 2 (Analyze).} Fit the analysis model to each completed dataset to obtain $\hat\beta^{(p)}$, and average these estimates to obtain $\hat\beta$. Evaluate $U_i^{(p)}$ and its derivative at $\hat\beta^{(p)}$, assigning zero contributions to subjects excluded by the analysis subset. Set $\widehat\tau$ to the negative average derivative over all $n$ subjects and $m$ imputations.}
\AlgState{\textbf{Step 3 (Components).} For each normal or logistic imputation model, calculate $S_{r,\mathrm{mis},i}^{(p)}$ and $d_{r,i}^{(p)}$ at the recorded draw $\psi_r^{(p)}$. Calculate its columns of $\widehat\kappa$ from the products of analysis and imputation scores within each completed dataset, as in Section~\ref{extension-1}.}
\AlgState{For PMM, obtain $\Sigma^{(p)}$ from the matching regression and calculate $p_{ij}^{(p)}$ and $\dot p_{ij}^{(p)}$ using \eqref{pmm-probability}--\eqref{pmm-probability-derivative}. Calculate $G_{ij}^{(p)}$ for the specified analysis and obtain $\widehat\kappa_{\mathrm{PMM}}$ from \eqref{pmm-kappa}. Calculate $d_{\mathrm{PMM},i}^{(p)}$ from the matching regression. Arrange all parameter-estimation contributions in $d_i^{(p)}$ in the same order as the columns of $\widehat\kappa$.}
\AlgState{\textbf{Step 4 (Assemble).} Average $d_i^{(p)}d_i^{(p)\top}$ over subjects and imputations to obtain $\widehat\Lambda$. For parametric imputation, calculate $\widehat\Omega$ and $\widehat C$ using $\bar U_i$. With PMM, first sum the scores by recorded donor within each completed dataset, then average across imputations and calculate $\widehat\Omega=\widehat\Omega_{\mathrm{PMM}}$ and $\widehat C=\widehat C_{\mathrm{PMM}}$ (Section~\ref{donor-source-analysis-scores}). Set $\widehat\Delta=\widehat\Omega+\widehat\kappa\widehat\Lambda\widehat\kappa^\top+\widehat C$.}
\AlgReturn{$\hat\beta$ and $\widehat\Gamma=n^{-1}\widehat\tau^{-1}\widehat\Delta\,\bigl(\widehat\tau^{-1}\bigr)^{\top}$.}
\end{algorithm}

For an untransformed PMM response in \texttt{lm()}, \texttt{pool\_rw()} calculates the PMM cross-moment automatically. For the GS binomial analysis, first call \texttt{pmm\_kappa\_binomial()} with the inclusion threshold, then pass the result to \texttt{pool\_rw()} using the \texttt{pmm\_kappa} argument.

\section{Experiments with simulated data}
\label{simulation-studies}
\subsection{Estimators and evaluation metrics}
\label{sim-estimators}
We compare RW with RR for variance estimation using both parametric imputation and PMM. For chained parametric MICE, RW-MICE denotes the estimator in Section~\ref{extension-1} and RR-MICE denotes Rubin's rules applied to the same imputations. For PMM, RW-PMM denotes the estimator in Section~\ref{extension-2} and RR-PMM denotes Rubin's rules applied to the same PMM imputations. Within each RW/RR pair, the completed datasets and point estimates are identical; only the variance estimator differs. Each simulation cell contains 2500 independent replications. For the \texttt{mice} analyses, each of the \(m\) imputation chains is run for five iterations.

We report the mean estimate, bias relative to the complete-data population parameter \(\beta_0\), empirical standard deviation, mean estimated standard error, coverage of \(\beta^\ast\), and coverage of \(\beta_0\). We approximate coverage of \(\beta^\ast\) by replacing it with the mean MI estimate across the 2500 replications in each cell. This substitution is approximate because the mean estimates the finite-sample expectation of \(\hat\beta\), which may differ from its probability limit. Both RW and RR intervals use \(\hat\beta\pm1.96\widehat{SE}\). Figure error bars are nominal 95\% binomial Monte Carlo intervals; those for coverage of \(\beta^\ast\) are approximate because the reference value is estimated from the same replications.
\subsection{Missing-data benchmark}
\label{simulation-rw-setting}
The missing-data benchmark follows the original simulation structure used by RW: a hypothetical example involving children at a daycare center.  Each subject has a group indicator \(A\), an age covariate \(X\), and a continuous outcome \(Z\). We use \(A=0\) for infants and \(A=1\) for toddlers, with \(P(A=1)=1/3\).  Infants have \(X\sim U(0.1,0.8)\) and toddlers have \(X\sim U(0.8,2.0)\).  In the main variants,
\[
  Z=X+\varepsilon,\qquad
  \varepsilon\mid X,A\sim N\{0,X^{\eta_A}\}.
\]
The outcome is fully observed for infants.  Among toddlers, \(Z\) is missing with probability 0.60 except in the outcome-dependent observation variant, where \(V=1\) denotes that \(Z\) is observed.  The completed-data analysis is the no-intercept least-squares estimating equation for the coefficient of \(X\) in \(E(Z\mid X)=\beta X\).  In one variant the analysis is restricted to toddlers; otherwise all subjects are included.  We generated data with both \(n=150\), the sample size used in the original RW simulations, and \(n=1000\). The six scenarios are:
\[
\begin{array}{ll}
\text{S1:}  & \eta_0=\eta_1=0,\ \text{analysis restricted to toddlers};\\
\text{S2a:} & \eta_0=\eta_1=1,\ \text{all-subject analysis};\\
\text{S2b:} & \eta_0=\eta_1=-1,\ \text{all-subject analysis};\\
\text{S2c:} & \eta_0=2,\ \eta_1=1,\ \text{all-subject analysis};\\
\text{S3a:} & \eta_0=\eta_1=0,\ P(V=1\mid Z,A=1)=\expit(2-0.5Z);\\
\text{S3b:} & Z=X+0.5X^2+\varepsilon,\ \eta_0=\eta_1=0.
\end{array}
\]
S1 is the subgroup analysis case. S2a--S2c use the same linear mean model with different conditional variances. S3a and S3b extend the original RW settings by changing the observation mechanism and the outcome mean, respectively. The complete-data population slope is \(\beta_0=1\) except in S3b, where the no-intercept linear analysis targets \(1+0.5E(X^3)/E(X^2)\approx1.690\).

The imputation model for \(Z\) uses \(X\) and \(A\) as predictors. We compare normal imputation and PMM with \(k=5\), using \(m=20\) at \(n=150\) and \(m=50\) at \(n=1000\).

Tables~\ref{tab:rw-benchmark-n150} and \ref{tab:rw-benchmark-n1000} report the results. At \(n=1000\), coverage of \(\beta^\ast\) ranges from 0.946 to 0.955 for RW-MICE and from 0.944 to 0.950 for RW-PMM. At \(n=150\), the corresponding ranges are 0.903--0.951 and 0.909--0.933, indicating undercoverage in several small-sample scenarios. RR is anti-conservative in S2a and S2c and conservative in S2b under both imputation methods.

\begin{table}[ht]
\centering
\scriptsize
\setlength{\tabcolsep}{3.5pt}
\begin{tabular}{llrrrrrr}
\toprule
Scenario & \makecell{Imputation\\model} & \makecell{Mean\\estimate} & Bias &
\makecell{Empirical\\SD} & \makecell{Mean estimated SE\\RW/RR} &
\makecell{Coverage of $\beta^\ast$\\RW/RR} & \makecell{Coverage of $\beta_0$\\RW/RR}\\
\midrule
S1 & Parametric MICE & 0.999 & -0.001 & 0.162 & 0.147 / 0.161 & 0.903 / 0.938 & 0.903 / 0.937 \\
S1 & PMM ($k=5$) & 0.980 & -0.020 & 0.165 & 0.151 / 0.153 & 0.912 / 0.916 & 0.909 / 0.912 \\
\addlinespace[2pt]
S2a & Parametric MICE & 0.999 & -0.001 & 0.160 & 0.147 / 0.104 & 0.911 / 0.795 & 0.911 / 0.795 \\
S2a & PMM ($k=5$) & 0.992 & -0.008 & 0.165 & 0.153 / 0.120 & 0.921 / 0.832 & 0.921 / 0.832 \\
\addlinespace[2pt]
S2b & Parametric MICE & 1.001 & 0.001 & 0.128 & 0.127 / 0.218 & 0.944 / 0.999 & 0.944 / 0.999 \\
S2b & PMM ($k=5$) & 0.967 & -0.033 & 0.126 & 0.119 / 0.183 & 0.929 / 0.992 & 0.924 / 0.992 \\
\addlinespace[2pt]
S2c & Parametric MICE & 0.999 & -0.001 & 0.159 & 0.145 / 0.088 & 0.905 / 0.732 & 0.905 / 0.732 \\
S2c & PMM ($k=5$) & 0.993 & -0.007 & 0.166 & 0.152 / 0.110 & 0.909 / 0.800 & 0.910 / 0.800 \\
\addlinespace[2pt]
S3a & Parametric MICE & 0.942 & -0.058 & 0.097 & 0.096 / 0.100 & 0.951 / 0.960 & 0.902 / 0.915 \\
S3a & PMM ($k=5$) & 0.936 & -0.064 & 0.100 & 0.096 / 0.098 & 0.933 / 0.942 & 0.881 / 0.891 \\
\addlinespace[2pt]
S3b & Parametric MICE & 1.670 & -0.020 & 0.142 & 0.132 / 0.136 & 0.916 / 0.931 & 0.916 / 0.928 \\
S3b & PMM ($k=5$) & 1.664 & -0.026 & 0.153 & 0.136 / 0.139 & 0.915 / 0.916 & 0.908 / 0.910 \\
\bottomrule
\end{tabular}
\caption{Missing-data benchmark at \(n=150\) and \(m=20\), with 2500 Monte Carlo replications per scenario.  Bias is the Monte Carlo mean estimate minus the complete-data population target \(\beta_0\). Coverage of \(\beta^\ast\) is approximated using the mean estimate in the same simulation cell.  RW/RR pairs refer to RW-MICE/RR-MICE for parametric MICE rows and RW-PMM/RR-PMM for PMM rows.}
\label{tab:rw-benchmark-n150}
\end{table}

\begin{table}[ht]
\centering
\scriptsize
\setlength{\tabcolsep}{3.5pt}
\begin{tabular}{llrrrrrr}
\toprule
Scenario & \makecell{Imputation\\model} & \makecell{Mean\\estimate} & Bias &
\makecell{Empirical\\SD} & \makecell{Mean estimated SE\\RW/RR} &
\makecell{Coverage of $\beta^\ast$\\RW/RR} & \makecell{Coverage of $\beta_0$\\RW/RR}\\
\midrule
S1 & Parametric MICE & 0.999 & -0.001 & 0.060 & 0.059 / 0.060 & 0.952 / 0.952 & 0.950 / 0.953 \\
S1 & PMM ($k=5$) & 1.000 & 0.000 & 0.063 & 0.061 / 0.061 & 0.947 / 0.937 & 0.947 / 0.937 \\
\addlinespace[2pt]
S2a & Parametric MICE & 0.999 & -0.001 & 0.061 & 0.060 / 0.040 & 0.946 / 0.799 & 0.948 / 0.800 \\
S2a & PMM ($k=5$) & 0.997 & -0.003 & 0.064 & 0.063 / 0.049 & 0.947 / 0.856 & 0.945 / 0.854 \\
\addlinespace[2pt]
S2b & Parametric MICE & 0.999 & -0.001 & 0.046 & 0.046 / 0.082 & 0.955 / 0.999 & 0.954 / 0.999 \\
S2b & PMM ($k=5$) & 0.996 & -0.004 & 0.047 & 0.046 / 0.073 & 0.948 / 0.996 & 0.949 / 0.995 \\
\addlinespace[2pt]
S2c & Parametric MICE & 0.999 & -0.001 & 0.060 & 0.059 / 0.034 & 0.946 / 0.718 & 0.946 / 0.721 \\
S2c & PMM ($k=5$) & 1.000 & 0.000 & 0.063 & 0.062 / 0.046 & 0.945 / 0.834 & 0.944 / 0.833 \\
\addlinespace[2pt]
S3a & Parametric MICE & 0.939 & -0.061 & 0.038 & 0.038 / 0.038 & 0.947 / 0.951 & 0.636 / 0.645 \\
S3a & PMM ($k=5$) & 0.939 & -0.061 & 0.039 & 0.038 / 0.038 & 0.950 / 0.952 & 0.633 / 0.638 \\
\addlinespace[2pt]
S3b & Parametric MICE & 1.672 & -0.018 & 0.053 & 0.052 / 0.052 & 0.948 / 0.943 & 0.932 / 0.927 \\
S3b & PMM ($k=5$) & 1.688 & -0.002 & 0.054 & 0.054 / 0.057 & 0.944 / 0.954 & 0.943 / 0.955 \\
\bottomrule
\end{tabular}
\caption{Missing-data benchmark at \(n=1000\) and \(m=50\), with 2500 Monte Carlo replications per scenario.  Columns are defined as in Table~\ref{tab:rw-benchmark-n150}.}
\label{tab:rw-benchmark-n1000}
\end{table}

\FloatBarrier
Coverage of \(\beta_0\) need not follow coverage of \(\beta^\ast\). In S3a at \(n=1000\), the mean estimate is 0.939 under both imputation methods, whereas \(\beta_0=1\); RW coverage of \(\beta_0\) is 0.636 for parametric imputation and 0.633 for PMM. In S3b, the PMM mean estimate of 1.688 is close to its complete-data population target of 1.690, and its coverage of \(\beta_0\) is 0.943. Thus bias can reduce coverage of \(\beta_0\) even when coverage of \(\beta^\ast\) is close to nominal. In the GS setting below, the imputed variable also determines eligibility for analysis.
\clearpage
\subsection{Measurement-error validation setting}
\label{simulation-gs-setting}
The second simulation setting follows the GS electronic health record validation example \citep{giganti2020multiple}. The analysis estimates the association between a continuous exposure \(A\) and a binary outcome \(D\) using logistic regression of \(D\) on \(A\), restricted to \(A>2\). Covariates \(X_1,X_2\) and error-prone versions \(A^\ast,D^\ast\) of $A$ and $D$ are observed for everyone. A simple random validation sample provides the values of \(A,D\); both are missing for the remaining subjects. We use \(n=2000\) with 500 validated records and \(n=4000\) with 1000 validated records. The complete-data population slope is \(\beta_0\approx0.88538\), obtained by numerical integration of the population logistic score under the data-generating mechanism. This target is the same for all sample sizes and imputation settings.

The parametric imputation model for \(A\) is normal, with predictors \(X_1,X_2,A^\ast,D^\ast\). The logistic model for \(D\) uses these predictors and the completed \(A\). Each iteration imputes \(A\) first, then \(D\) using the updated \(A\). For PMM, \texttt{mice} imputes \(A\) by donor matching; the \(D\) model and update order are unchanged. The probabilities \(p_{ij}\) and expected scores \(G_{ij}\) in Section~\ref{extension-2} are used only to calculate the PMM variance components. In the GS simulations, the conditional expectations used to calculate the PMM cross-moment were evaluated by numerical integration.

Tables~\ref{tab:gs-validation-n2000} and \ref{tab:gs-validation-n4000} report parametric imputation and PMM with \(k=5,10\), for \(m=5,25,50,100\). Results for \(k=20,30\) are in Tables~\ref{tab:pmm-k-sensitivity-supp} and \ref{tab:pmm-k4000-supp}. Figures~\ref{fig:gs-pmm-current-n2000-supp} and \ref{fig:gs-pmm-current-n4000-supp} show the PMM coverage results for all four donor pool sizes. Additional parametric simulations at \(n=6000,8000\), with \(n/4\) validated records, are included in Figure~\ref{fig:gs-parametric-full-supp}.

\begin{table}[!htbp]
\centering
\scriptsize
\setlength{\tabcolsep}{2.5pt}
\begin{tabular}{llrrrrrr}
\toprule
$m$ & \makecell{Imputation\\model} & \makecell{Mean\\estimate} & Bias &
\makecell{Empirical\\SD} & \makecell{Mean estimated SE\\RW/RR} &
\makecell{Coverage of $\beta^\ast$\\RW/RR} & \makecell{Coverage of $\beta_0$\\RW/RR}\\
\midrule
5 & Parametric MICE & 0.856 & -0.029 & 0.117 & 0.114 / 0.166 & 0.944 / 0.990 & 0.928 / 0.984 \\
5 & PMM ($k=5$) & 0.856 & -0.030 & 0.133 & 0.131 / 0.168 & 0.946 / 0.982 & 0.936 / 0.971 \\
5 & PMM ($k=10$) & 0.848 & -0.037 & 0.133 & 0.131 / 0.168 & 0.948 / 0.977 & 0.929 / 0.967 \\
\addlinespace[2pt]
25 & Parametric MICE & 0.852 & -0.033 & 0.110 & 0.107 / 0.163 & 0.940 / 0.998 & 0.924 / 0.993 \\
25 & PMM ($k=5$) & 0.857 & -0.028 & 0.127 & 0.126 / 0.165 & 0.949 / 0.987 & 0.940 / 0.983 \\
25 & PMM ($k=10$) & 0.848 & -0.038 & 0.126 & 0.125 / 0.166 & 0.950 / 0.988 & 0.928 / 0.984 \\
\addlinespace[2pt]
50 & Parametric MICE & 0.852 & -0.034 & 0.107 & 0.106 / 0.163 & 0.951 / 0.997 & 0.935 / 0.994 \\
50 & PMM ($k=5$) & 0.856 & -0.029 & 0.122 & 0.125 / 0.164 & 0.950 / 0.993 & 0.946 / 0.985 \\
50 & PMM ($k=10$) & 0.848 & -0.037 & 0.123 & 0.124 / 0.165 & 0.952 / 0.992 & 0.938 / 0.983 \\
\addlinespace[2pt]
100 & Parametric MICE & 0.853 & -0.032 & 0.107 & 0.106 / 0.162 & 0.947 / 0.996 & 0.926 / 0.995 \\
100 & PMM ($k=5$) & 0.855 & -0.031 & 0.121 & 0.124 / 0.164 & 0.953 / 0.994 & 0.945 / 0.988 \\
100 & PMM ($k=10$) & 0.848 & -0.037 & 0.123 & 0.124 / 0.165 & 0.951 / 0.991 & 0.931 / 0.984 \\
\bottomrule
\end{tabular}
\caption{Measurement-error validation setting at \(n=2000\), with 2500 Monte Carlo replications per cell.  RW/RR pairs refer to RW-MICE/RR-MICE for parametric MICE rows and RW-PMM/RR-PMM for PMM rows.}
\label{tab:gs-validation-n2000}
\end{table}

\begin{table}[!htbp]
\centering
\scriptsize
\setlength{\tabcolsep}{2.5pt}
\begin{tabular}{llrrrrrr}
\toprule
$m$ & \makecell{Imputation\\model} & \makecell{Mean\\estimate} & Bias &
\makecell{Empirical\\SD} & \makecell{Mean estimated SE\\RW/RR} &
\makecell{Coverage of $\beta^\ast$\\RW/RR} & \makecell{Coverage of $\beta_0$\\RW/RR}\\
\midrule
5 & Parametric MICE & 0.867 & -0.019 & 0.084 & 0.082 / 0.117 & 0.946 / 0.987 & 0.937 / 0.984 \\
5 & PMM ($k=5$) & 0.872 & -0.014 & 0.095 & 0.095 / 0.120 & 0.946 / 0.976 & 0.940 / 0.972 \\
5 & PMM ($k=10$) & 0.869 & -0.016 & 0.096 & 0.094 / 0.119 & 0.944 / 0.979 & 0.937 / 0.974 \\
\addlinespace[2pt]
25 & Parametric MICE & 0.867 & -0.018 & 0.077 & 0.078 / 0.115 & 0.954 / 0.995 & 0.947 / 0.994 \\
25 & PMM ($k=5$) & 0.870 & -0.015 & 0.089 & 0.090 / 0.115 & 0.956 / 0.988 & 0.952 / 0.985 \\
25 & PMM ($k=10$) & 0.869 & -0.017 & 0.091 & 0.090 / 0.116 & 0.946 / 0.988 & 0.939 / 0.982 \\
\addlinespace[2pt]
50 & Parametric MICE & 0.867 & -0.018 & 0.077 & 0.077 / 0.114 & 0.950 / 0.996 & 0.939 / 0.994 \\
50 & PMM ($k=5$) & 0.872 & -0.014 & 0.089 & 0.090 / 0.115 & 0.949 / 0.988 & 0.942 / 0.988 \\
50 & PMM ($k=10$) & 0.870 & -0.015 & 0.089 & 0.089 / 0.115 & 0.948 / 0.990 & 0.944 / 0.988 \\
\addlinespace[2pt]
100 & Parametric MICE & 0.866 & -0.019 & 0.079 & 0.077 / 0.114 & 0.948 / 0.995 & 0.934 / 0.992 \\
100 & PMM ($k=5$) & 0.870 & -0.015 & 0.088 & 0.090 / 0.114 & 0.953 / 0.989 & 0.947 / 0.986 \\
100 & PMM ($k=10$) & 0.872 & -0.014 & 0.090 & 0.089 / 0.115 & 0.948 / 0.988 & 0.941 / 0.987 \\
\bottomrule
\end{tabular}
\caption{Measurement-error validation setting at \(n=4000\), with 2500 Monte Carlo replications per cell.  Columns are defined as in Table~\ref{tab:gs-validation-n2000}.}
\label{tab:gs-validation-n4000}
\end{table}

\begin{figure}[ht]
\centering
\includegraphics[width=\textwidth]{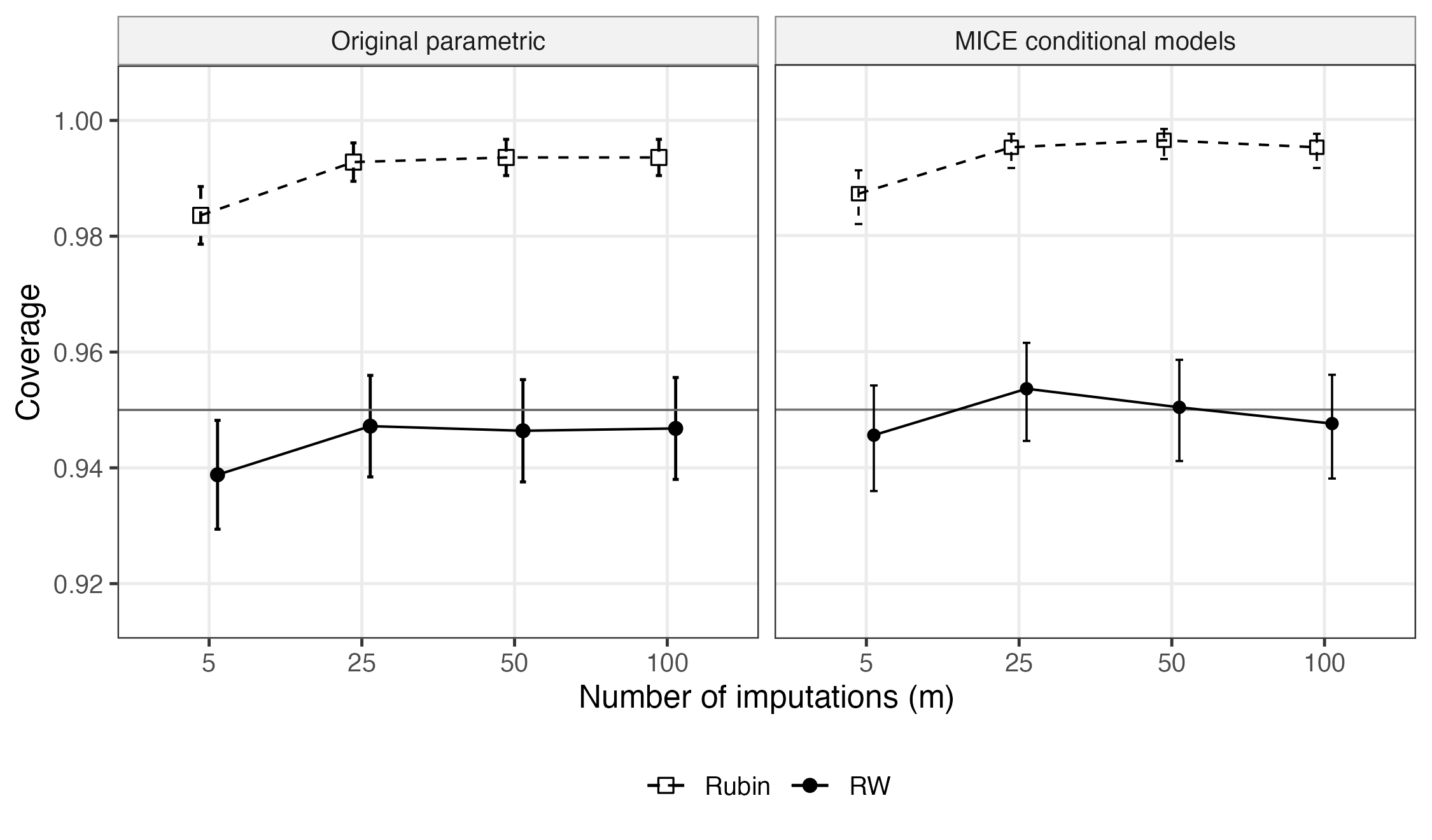}
\caption{Coverage of \(\beta^\ast\) in the Giganti--Shepherd measurement-error setting with parametric imputation at \(n=4000\). The left panel shows the original Giganti--Shepherd parametric implementation, and the right panel shows the chained conditional MICE implementation. Each panel compares Rubin's rules with RW variance estimation.}
\label{fig:gs-parametric}
\end{figure}

\FloatBarrier
For parametric imputation, RW coverage of \(\beta^\ast\) is 0.940--0.951 at \(n=2000\) and 0.946--0.954 at \(n=4000\). RR mean standard errors exceed the empirical standard deviations by approximately 39--52\% across these cells. Figure~\ref{fig:gs-parametric} compares coverage of \(\beta^\ast\) under the original GS parametric implementation and MICE conditional models. Both coverage measures for the MICE simulations are shown in Figure~\ref{fig:gs-parametric-full-supp}.

For PMM, RW coverage of \(\beta^\ast\) is 0.932--0.953 at \(n=2000\) and 0.938--0.956 at \(n=4000\), across all 16 combinations of \(m\) and \(k\) at each sample size. RR coverage of \(\beta^\ast\) is 0.977--0.994 and 0.973--0.990, respectively. Coverage of \(\beta_0\) is lower for RW-PMM: 0.884--0.946 at \(n=2000\) and 0.920--0.952 at \(n=4000\). Larger donor pools are associated with a more negative bias in the PMM point estimate, while coverage of \(\beta^\ast\) changes less. For example, at \(n=4000,m=25\), increasing \(k\) from 5 to 30 changes the bias from \(-0.015\) to \(-0.032\), coverage of \(\beta^\ast\) from 0.956 to 0.942, and coverage of \(\beta_0\) from 0.952 to 0.924. This comparison concerns the unchanged \texttt{mice} point estimator; the variance correction does not remove its bias.

\clearpage

\section{Discussion}\label{discussion}
We apply the estimating-equation framework of \citet{robins2000inference} to chained conditional imputation and PMM. For chained imputation, the RW components are calculated from the conditional-model scores and influence contributions. For PMM, we approximate the cross-moment using donor probabilities and expected analysis scores, and sum analysis scores by recorded donor to account for donor reuse. Both procedures are implemented in the R package \texttt{rw} and retain the completed-data point estimates used by Rubin's rules. This should facilitate the uptake of the RW variance estimator, which is valid in a wider variety of settings than the RR variance estimator. By extending to PMM, we allow the default imputation method from the \texttt{mice} package to be used with a more robust variance estimator.

The failure of RR under uncongeniality has been understood since \citet{meng1994multiple} and formalized for estimating-equation analyses by \citet{wang1998large}. Our simulations include settings in which RR is either anti-conservative or conservative. In the RW benchmark at \(n=1000\), the proposed estimators give close to nominal coverage of \(\beta^\ast\), whereas undercoverage remains in several scenarios at \(n=150\). In the GS validation setting, motivated by EHR analyses with imputed inclusion criteria \citep{shen2026missingness}, RR gives conservative intervals and RW generally brings coverage of \(\beta^\ast\) closer to nominal. Coverage of \(\beta_0\) can nevertheless remain below nominal, particularly with larger PMM donor pools, because changing the variance estimator does not remove bias in the point estimate.

Our approach is appealing both theoretically and computationally. The bootstrap offers an alternative to analytic variance estimation. Resampling the incomplete records and repeating imputation and analysis within each bootstrap sample can provide valid inference under uncongeniality under the conditions for the chosen bootstrap procedure \citep{schomaker2018bootstrap, bartlett2020bootstrap}. The RW calculation avoids repeating the full imputation procedure for each bootstrap sample.

Our approach has several limitations. The parametric justification requires large \(m\) and conditional-model components that recover the derivative of the expected analysis score and the influence function of the full imputation-parameter estimate. The sequential normal and logistic fits in the GS setting satisfy these conditions; more general MICE specifications require verification. Additional variation from the parameter draws can matter when \(m\) is small. For PMM, the appendix derives the limit of the cross-moment with known matching coefficients and the covariance induced by shared donor values. These calculations support the proposed approximation, but consistency of the complete variance estimator with fitted PMM models remains to be established. Its accuracy depends in part on the conditional means of the summed scores and on estimation of the matching model.

The PMM calculation requires donor indices recorded during imputation. The package saves these indices without changing the \texttt{mice} draws. The present construction imputes one continuous variable by PMM and any remaining incomplete variables using parametric conditional models. Extending it to multiple PMM variables would require accounting for dependence from donor reuse across variables. Extensions beyond linear and logistic regression would require the corresponding analysis scores and expected scores \(G_{ij}\).

\section*{Acknowledgments}
This manuscript was supported in part by the U.S. National Institutes of Health (NIH) grant R37AI131771 (BDW).

\section*{Code availability}
We provide code to reproduce both simulations and data analyses on GitHub at \url{https://github.com/LucyMcGowan/rw}.

\section*{Disclosures}
BDW has worked on grants awarded to Kaiser Permanente Washington Health Research Institute (KPWHRI) by Eli Lilly.

\appendix
\section{Variance Derivations}\label{appendix-theory}
We now derive the parametric large-\(m\) variance and the probabilities, cross-moment, and covariance calculations used for RW-PMM. Write \(P_n v_i=n^{-1}\sum_i v_i\), and treat estimating functions as column vectors.

\subsection{Parametric imputation}\label{appendix mice}
\paragraph{Fitted imputation parameters.}
Let \(O_i\) denote the observed record and \(q_\psi(z_{\mathrm{mis}}\mid O_i)\) the conditional imputation density. Define
\[
h_i(\psi,\beta)
=\int U(O_i,z_{\mathrm{mis}};\beta)
q_\psi(z_{\mathrm{mis}}\mid O_i)\,dz_{\mathrm{mis}},
\qquad h_i=h_i(\psi^\ast,\beta^\ast),
\]
where \(\psi^\ast\) is the limiting fitted imputation parameter and \(\beta^\ast\) solves \(E\{h_i(\psi^\ast,\beta)\}=0\). Integration includes summation for discrete missing variables. For a fully observed record, \(h_i\) is its analysis score.

Suppose \(\hat\psi\) solves \(P_n S_{\mathrm{obs},i}(\hat\psi)=0\). With \(H=E\{\partial S_{\mathrm{obs},i}(\psi^\ast)/\partial\psi^\top\}\) nonsingular, a Taylor expansion gives
\[
0=P_n S_{\mathrm{obs},i}(\psi^\ast)
+H(\hat\psi-\psi^\ast)+o_p(n^{-1/2}).
\]
Consequently,
\[
\sqrt n(\hat\psi-\psi^\ast)
=\frac{1}{\sqrt n}\sum_{i=1}^n d_i+o_p(1),
\qquad d_i=-H^{-1}S_{\mathrm{obs},i}(\psi^\ast),
\qquad E(d_i)=0.
\]
Here and below, assume the local smoothness, stochastic equicontinuity, and moment conditions for the RW expansion, including a common nonsingular limiting analysis-score derivative and the moment bounds needed for covariance convergence \citep{robins2000inference}. Records are independent, and missing values are generated independently across records conditional on the observed sample and the imputation parameters.

\paragraph{Sensitivity of the mean analysis score.}
For parameter-independent support and differentiation under the integral, let \(S_{\mathrm{mis},i}=\partial\log q_\psi/\partial\psi\). Then
\begin{align*}
\frac{\partial h_i(\psi,\beta)}{\partial\psi^\top}
&=\int U(O_i,z_{\mathrm{mis}};\beta)
\frac{\partial q_\psi(z_{\mathrm{mis}}\mid O_i)}{\partial\psi^\top}
\,dz_{\mathrm{mis}}\\
&=E_\psi\{U_i(\beta)S_{\mathrm{mis},i}^\top\mid O_i\}.
\end{align*}
Taking expectation over observed records and evaluating at the limiting parameters gives
\[
\kappa
=E\left\{\frac{\partial h_i(\psi,\beta^\ast)}
{\partial\psi^\top}\bigg|_{\psi=\psi^\ast}\right\}
=E\{U_i S_{\mathrm{mis},i}^\top\}.
\]
Thus \(\kappa\) describes the first-order change in the mean analysis equation caused by a change in the fitted imputation parameters.

\paragraph{Completed-data estimates and their average.}
Write \(U_i^{(p)}(\beta)\) for the score in completed dataset \(p\). With \(\tau=-E\{\partial U_i^{(p)}(\beta^\ast)/\partial\beta^\top\}\), a Taylor expansion of \(P_n U_i^{(p)}(\hat\beta^{(p)})=0\) gives
\[
0=P_nU_i^{(p)}(\beta^\ast)
-\tau(\hat\beta^{(p)}-\beta^\ast)+o_p(n^{-1/2}).
\]
For fixed \(m\), averaging these expansions yields
\[
\sqrt n(\hat\beta-\beta^\ast)
=\tau^{-1}\frac{1}{m}\sum_{p=1}^m
\sqrt n\,P_nU_i^{(p)}(\beta^\ast)+o_p(1),
\qquad \hat\beta=\frac{1}{m}\sum_p\hat\beta^{(p)}.
\]
This expansion applies to the average of the completed-data estimates used in the implementation.

To separate sampling and imputation variation, suppose imputation \(p\) uses \(\psi^{(p)}=\hat\psi+n^{-1/2}Z_{np}\). Conditional on the observed sample, assume the parameter draws are independent and centered, with \(\operatorname{Var}(Z_{np}\mid O_{1:n})\to_p Q\). The variation from generating missing values at a given parameter draw is
\[
M_{np}=\frac{1}{\sqrt n}\sum_i
\{U_i^{(p)}(\beta^\ast)-h_i(\psi^{(p)},\beta^\ast)\}.
\]
Its conditional mean given the observed sample and parameter draws is zero. Conditional independence and moment convergence give
\[
\operatorname{Var}(M_{np}\mid O_{1:n},\psi^{(p)})\to_p V_{\mathrm{mis}},
\qquad V_{\mathrm{mis}}=E[\operatorname{Var}_{\psi^\ast}\{U_i(\beta^\ast)\mid O_i\}].
\]
Expanding the conditional mean now gives
\begin{align*}
\sqrt n\,P_n U_i^{(p)}(\beta^\ast)
&=\sqrt n\,P_n h_i(\psi^{(p)},\beta^\ast)+M_{np}\\
&=\frac{1}{\sqrt n}\sum_i h_i
+\kappa\sqrt n(\hat\psi-\psi^\ast)
+\kappa Z_{np}+M_{np}+o_p(1).
\end{align*}
Substituting the fitted-parameter expansion and averaging over \(p\) therefore yields
\[
\sqrt n(\hat\beta-\beta^\ast)
=\tau^{-1}\left\{
\frac{1}{\sqrt n}\sum_i(h_i+\kappa d_i)
+\bar M_n+\kappa\bar Z_n\right\}+o_p(1),
\]
where \(\bar M_n=m^{-1}\sum_p M_{np}\) and \(\bar Z_n=m^{-1}\sum_p Z_{np}\). With independent imputation randomizations, their covariance contributions are \(V_{\mathrm{mis}}/m\) and \(\kappa Q\kappa^\top/m\). Conditional centering makes both uncorrelated with observed-sample contributions, and makes \(\bar M_n\) uncorrelated with \(\bar Z_n\).

\paragraph{The three variance terms.}
Since \(E(h_i)=E(d_i)=0\), direct expansion gives
\begin{align*}
\Delta_\infty
&=E\{(h_i+\kappa d_i)(h_i+\kappa d_i)^\top\}\\
&=E(h_i h_i^\top)+\kappa E(d_i d_i^\top)\kappa^\top
+\kappa E(d_i h_i^\top)+E(h_i d_i^\top)\kappa^\top\\
&=\Omega_\infty+\kappa\Lambda\kappa^\top+C_\infty.
\end{align*}
Here \(\Omega_\infty=E(h_i h_i^\top)\) is the covariance of the conditional mean analysis scores, \(\Lambda=E(d_i d_i^\top)\) is the covariance of the imputation-parameter influence function, and \(C_\infty\) contains the cross-covariance terms. Under the assumptions above, the fixed-\(m\) asymptotic covariance on the root-\(n\) scale is
\[
V_m=\tau^{-1}\left\{\Delta_\infty+
\frac{V_{\mathrm{mis}}+\kappa Q\kappa^\top}{m}\right\}(\tau^{-1})^\top.
\]
The matrix \(Q\) is the parameter-draw covariance, not necessarily the sampling covariance \(\Lambda\). In particular, parameter draws contribute \(n^{-1}m^{-1}\tau^{-1}\kappa Q\kappa^\top(\tau^{-1})^\top\) to the coefficient covariance. Both imputation Monte Carlo terms vanish as \(m\) increases, leaving
\[
V_\infty=\tau^{-1}\Delta_\infty(\tau^{-1})^\top.
\]
This conclusion can be read in the sequential limit, first \(n\to\infty\), then \(m\to\infty\). A joint limit with \(m=m_n\to\infty\) additionally requires the averaged Taylor remainders to remain \(o_p(1)\) and the moment bounds to hold along that sequence.

In the same large-\(m\) limit, suppose the empirical components in Section~\ref{extension-1} satisfy
\[
(\widehat\tau,\widehat\Omega,\widehat\kappa,
\widehat\Lambda,\widehat C)
\ \longrightarrow_p\
(\tau,\Omega_\infty,\kappa,\Lambda,C_\infty).
\]
These limits require the recorded contributions to estimate the influence function \(d_i\) of \(\hat\psi\), the score cross-products to estimate \(\kappa\), and laws of large numbers for the corresponding products. Continuity of matrix inversion at nonsingular \(\tau\) then gives \(n\widehat\Gamma_{\mathrm{MICE}}\to_p V_\infty\). This large-\(m\) result is consistent with the infinite-imputation comparison in \citet{robins2000inference}. The additional parameter-draw term in \(V_m\) is not included in the implemented formula.

\paragraph{Sequential fitting in the GS setting.}
Let \(W\) contain the fully observed predictors and let \(R_i\) indicate validation of both \(A_i\) and \(D_i\). The conditional imputation density factors as
\[
q_\psi(a,d\mid W)=f_{\eta_A,\sigma_A^2}(a\mid W)g_\gamma(d\mid a,W),
\qquad \psi=(\eta_A^\top,\sigma_A^2,\gamma^\top)^\top,
\]
with a normal model for \(A\), residual variance \(\sigma_A^2\), and a logistic model for \(D\). Both fits use validation records, where their predictors are observed. The observed score is \(R_i(s_{A,i}^\top,s_{D,i}^\top)^\top\), with distinct parameter vectors in the two factors, so
\[
H=\begin{pmatrix}H_A&0\\0&H_D\end{pmatrix},
\qquad
d_i=\begin{pmatrix}-H_A^{-1}R_i s_{A,i}\\-H_D^{-1}R_i s_{D,i}\end{pmatrix}.
\]
Thus combining the contributions from the two fits gives the influence function of the full parameter estimate \(\hat\psi\), while \(\Lambda=E(d_i d_i^\top)\) retains cross-moments between the fits. The analysis restriction \(A>c_0\) is included in \(U\) and hence in the integral defining \(h_i\). For other chained procedures, the same derivation requires the recorded contributions to estimate the influence function of the full parameter vector and the score cross-products to estimate the derivative \(\kappa\).

\subsection{PMM variance calculation}\label{appendix pmm}
\paragraph{Donor probabilities.}
Suppress the imputation index and use the notation of Section~\ref{smooth-copied-donor}. Order the distinct donor predictions as \(a_{(1)}<\cdots<a_{(N)}\). Donor \((r)\) loses its place among the nearest \(k\) donors to a prediction on its left when \(k\) lower predictions become closer; the boundary is the midpoint of \(a_{(r-k)}\) and \(a_{(r)}\). The argument on the right is the same. Thus the donor is selected into the pool for \(\ell_{(r)}<t<u_{(r)}\), where
\[
\ell_{(r)}=\begin{cases}
-\infty,&r\leq k,\\
(a_{(r-k)}+a_{(r)})/2,&r>k,
\end{cases}
\qquad
u_{(r)}=\begin{cases}
(a_{(r)}+a_{(r+k)})/2,&r+k\leq N,\\
\infty,&r+k>N.
\end{cases}
\]
Let \(\mathcal N(t)\) denote this nearest-donor pool and \(T_i\sim N(t_i,s_i^2)\), with \(s_i>0\). Integrating its uniform donor selection gives
\begin{align*}
p_{ij}
&=\int\frac{\mathbf 1\{j\in\mathcal N(t)\}}{k}
\frac{1}{s_i}\phi\left(\frac{t-t_i}{s_i}\right)\,dt\\
&=\frac{1}{k}\int_{\ell_j}^{u_j}
\frac{1}{s_i}\phi\left(\frac{t-t_i}{s_i}\right)\,dt
=\frac{\Phi(z_{ij}^{u})-\Phi(z_{ij}^{\ell})}{k},
\end{align*}
where \(z_{ij}^{u}=(u_j-t_i)/s_i\) and \(z_{ij}^{\ell}=(\ell_j-t_i)/s_i\). Since \(\sum_j\mathbf 1\{j\in\mathcal N(t)\}=k\) away from ties, summing the integral over donors proves \(\sum_jp_{ij}=1\).

\paragraph{Probability derivatives.}
Shift both matching coefficient vectors by \(\delta\), keeping \(\Sigma\) fixed. Locally the donor ordering is unchanged, and
\[
\nabla_\delta t_i=x_i,\qquad
\dot\ell_{(r)}=(x_{(r-k)}+x_{(r)})/2,\qquad
\dot u_{(r)}=(x_{(r)}+x_{(r+k)})/2
\]
for finite endpoints. Applying the chain rule to each normal distribution function gives
\[
\dot p_{ij}
=\frac{\phi(z_{ij}^{u})(\dot u_j-x_i)
-\phi(z_{ij}^{\ell})(\dot\ell_j-x_i)}{k s_i}.
\]
Infinite endpoints contribute zero. Differentiating \(\sum_jp_{ij}=1\) proves \(\sum_j\dot p_{ij}=0\). An intercept shift moves every prediction equally, making each endpoint difference zero in that coordinate. Derivatives for the other coordinates can be nonzero.

\paragraph{Effect of normal averaging.}
For \(Z\sim N(0,1)\) and a twice differentiable scalar function \(g\) with bounded second derivative, Taylor's theorem with integral remainder gives
\[
g(t+s_iZ)-g(t)
=s_iZg'(t)+s_i^2 Z^2\int_0^1(1-v)g''(t+v s_iZ)\,dv.
\]
Taking expectation over \(Z\), using \(E(Z)=0\) and \(E(Z^2)=1\), yields
\[
\left|E\{g(t+s_iZ)\}-g(t)\right|
\leq \frac{s_i^2}{2}\sup_u|g''(u)|.
\]
For bounded matching predictors and \(\Sigma=O_p(n^{-1})\), \(s_i^2=x_i^\top\Sigma x_i=O_p(n^{-1})\). Under a uniform bound on \(g''\), the difference is therefore \(o_p(n^{-1/2})\). If \(g'''\) is bounded, applying the same argument to \(g'\) bounds the derivative error by \(s_i^2\sup|g'''|/2\), holding \(s_i\) fixed. These bounds apply to smooth population expectations. The finite-sample nearest-donor rule is a step function, and the normal averaging used in \(G_{ij}\) involves a residual variance that need not shrink with \(n\); neither is covered by these bounds.

\paragraph{Expected analysis scores.}
Let \(J_i^\dagger\) have probabilities \(p_{ij}\) and define \(G_{ij}=E(U_i^\dagger\mid J_i^\dagger=j,\text{fitted models and observed covariates})\). The expectation uses the fitted distributions of the donor value and any subsequent imputation specified in Section~\ref{extension-2}. With \(S_{ij}=\dot p_{ij}/p_{ij}\) for \(p_{ij}>0\), conditioning first on the donor index gives
\[
E(U_i^\dagger S_{iJ_i^\dagger}^\top)
=\sum_jp_{ij}G_{ij}S_{ij}^\top
=\sum_jG_{ij}\dot p_{ij}^\top.
\]
Averaging over recipients and imputations gives
\[
\widehat\kappa_{\mathrm{PMM}}
=\frac{1}{nm}\sum_{p=1}^m\sum_{i\in\mathcal M_A}
\sum_{j\in\mathcal O_A}G_{ij}^{(p)}\dot p_{ij}^{(p)\top},
\]
as in \eqref{pmm-kappa}. For each recipient, the sum over donors is the derivative of \(\sum_jp_{ij}(\delta)G_{ij}\) with \(G_{ij}\) fixed. Differentiating \(G_{ij}\) would additionally change the models for the donor value and subsequent imputations. Parametric columns retain their score cross-products, in the same parameter order as \(d_i^{(p)}\). The analysis scores from the completed data enter the remaining variance components.

With known matching and subsequent imputation parameters, \(G_{ij}\) can equal the conditional mean analysis score under the data-generating model. Assume that donor selection depends on predictors and independent matching randomization, not on donor residuals. For a linear response score, \(E(A_j\mid x_j,j\in\mathcal O_A)=a_j\) gives \(G_{ij}=z_i(a_j-z_i^\top\beta)/v\) directly, without assumptions of normality or constant residual variance. In the GS generating model, \(A_j\mid W_j\sim N(a_j,1)\) and \(D_i\mid A,W_i\) has logistic mean \(\rho_i(A)\). Define
\[
g_i(a)=\mathbf 1\{a>c_0\}z(a)\{\rho_i(a)-\pi_\beta(a)\}.
\]
For a missing recipient, averaging its binary outcome and then the donor value gives
\[
G_{ij}=E\{g_i(A_j)\mid W_i,W_j,J_i=j\}
=\int g_i(a)\phi(a-a_j)\,da.
\]
This expectation conditions on the donor's predictors and averages over its value. The normal integral includes the original analysis restriction. Equation~\eqref{pmm-binomial-score} substitutes fitted parameters and evaluates the integral by numerical quadrature.

\paragraph{An equivalent expression for the cross-moment.}
Write \(G_{i(r)}=G_i(a_{(r)})\), keeping the recipient's observed covariates and the models used in \(G_i\) fixed. Use the finite upper endpoints \(u_{(r)}\) and their derivatives \(\dot u_{(r)}\) defined above. For \(r=1,\ldots,N-k\), set
\[
\Delta_r=a_{(r+k)}-a_{(r)},\qquad K_s(u)=\phi(u/s)/s.
\]
The upper endpoint of donor \((r)\)'s interval is the lower endpoint of donor \((r+k)\)'s interval. Pairing their derivative terms therefore yields the exact finite-sample identity
\begin{equation}
\mathcal K_i:=\sum_jG_{ij}\dot p_{ij}^\top
=\frac{1}{k}\sum_{r=1}^{N-k}
\{G_i(a_{(r)})-G_i(a_{(r+k)})\}
K_{s_i}(u_{(r)}-t_i)(\dot u_{(r)}-x_i)^\top.
\label{pmm-neighbor-identity}
\end{equation}
Infinite endpoints contribute zero, so there are no remaining boundary terms. Equivalently, define
\[
\omega_{ir}=\frac{\Delta_r}{k}K_{s_i}(u_{(r)}-t_i),\qquad
Q_{ir}=\frac{G_i(a_{(r+k)})-G_i(a_{(r)})}{\Delta_r}.
\]
Then \(\mathcal K_i=\sum_r\omega_{ir}Q_{ir}(x_i-\dot u_{(r)})^\top\). The weights are nonnegative, although their finite-sample sum need not equal one. For nearby donors, the expected-score difference is small when \(G_i\) is smooth; this difference offsets the factor \(1/s_i\) in the probability derivative. Equation~\eqref{pmm-neighbor-identity} is an algebraic rearrangement of the same cross-moment.

\paragraph{Limit with known matching coefficients.}
Consider a fixed recipient with prediction \(t\) in the interior of a compact donor-prediction interval, with known matching coefficients and fixed \(k\). Let \(s_N\to0\) and let the maximum prediction spacing, including endpoint gaps, satisfy \(h_N/s_N\to_p0\). Assume that \(G_i\) has a bounded continuous derivative on the interval. Conditional on their predictions, assume that donor predictors are independent, have uniformly bounded second moments, and have conditional mean \(\mu_x(a)=E(x_j\mid a_j=a,j\in\mathcal O_A)\) continuous at \(t\). Then
\begin{equation}
\mathcal K_i\ \longrightarrow_p\
G_i'(t)\{x_i-\mu_x(t)\}^\top.
\label{pmm-matching-limit}
\end{equation}
To see this, partition the intervals \([a_{(r)},a_{(r+k)}]\) into the \(k\) sequences indexed by \(r\) modulo \(k\). Within each sequence, consecutive intervals meet at their endpoints. Midpoint integration of the normal kernel gives \(\sum_r\omega_{ir}=1+o_p(1)\): its error is of order \(h_N/s_N\), and the endpoint tails vanish because \(t\) is interior. The weights outside any fixed neighborhood of \(t\) tend to zero. The weighted average of the difference quotients \(Q_{ir}\) therefore tends to \(G_i'(t)\), and the weighted average of the conditional predictor means tends to \(\mu_x(t)\). Each donor predictor appears in at most two endpoint averages \(\dot u_{(r)}\). Consequently, the conditional variance of the centered weighted sum is bounded by a constant times
\[
\sum_r\omega_{ir}^2
\leq \max_r\omega_{ir}\sum_r\omega_{ir}
=O_p(h_N/s_N)=o_p(1).
\]
Substitution in \eqref{pmm-neighbor-identity} proves \eqref{pmm-matching-limit}. For independent predictions with density bounded away from zero on a compact interval, the maximum spacing is \(O_p(\log N/N)\). Indeed, a partition into intervals of length proportional to \(\log N/N\), an exponential bound for each empty-interval probability, and a union bound give this rate. Thus the spacing requirement is compatible with \(s_N\) of order \(N^{-1/2}\).

The limit has a population interpretation. Suppose \(a_j=x_j^\top\theta_0\) and the expected recipient score conditional on the donor predictors is \(G_i(a_j)\). This holds for the linear conditional-mean model above and for a response of the form \(A_j=a_j+\epsilon_j\) with a common residual distribution independent of the predictors. Perturb only the matching coefficient, keeping the response and subsequent imputation models fixed, and define
\[
M_i(\delta)=E\{G_i(x_j^\top\theta_0)\mid
x_j^\top(\theta_0+\delta)=x_i^\top(\theta_0+\delta),\ j\in\mathcal O_A\}.
\]
On the conditioning event, \(x_j^\top\theta_0=t+\delta^\top(x_i-x_j)\). Under continuity of the conditional predictor mean as \(\delta\to0\) and moment bounds allowing a conditional Taylor expansion,
\[
M_i(\delta)=G_i(t)+G_i'(t)\delta^\top\{x_i-\mu_x(t)\}
+o(\|\delta\|).
\]
Thus \eqref{pmm-matching-limit} equals \(\nabla_\delta M_i(0)\), the change in the expected analysis score when the matching coefficients change while the response model remains fixed. The result concerns one recipient with known matching coefficients. Extending it to all recipients when the coefficients and donor ordering are estimated from the same sample requires additional analysis.

\paragraph{Summing scores by donor.}
Use the recorded donor indices \(J_i^{(p)}\) from Section~\ref{extension-2}, setting \(J_i^{(p)}=i\) when the PMM variable is observed. The sets \(\mathcal I_j^{(p)}=\{i:J_i^{(p)}=j\}\) partition the records by donor, so
\[
\widetilde U_j^{(p)}(\beta)=\sum_{i\in\mathcal I_j^{(p)}}U_i^{(p)}(\beta),
\qquad
\sum_j\widetilde U_j^{(p)}(\beta)=\sum_i U_i^{(p)}(\beta).
\]
The identity holds for every \(\beta\), and therefore also for its derivative when the recorded donor assignments are fixed. Both the point estimate and the analysis derivative are preserved. Expanding the outer product gives
\[
\widetilde U_j^{(p)}\widetilde U_j^{(p)\top}
=\sum_{i\in\mathcal I_j^{(p)}}U_i^{(p)}U_i^{(p)\top}
+\sum_{\substack{i,\ell\in\mathcal I_j^{(p)}\\i\ne\ell}}
U_i^{(p)}U_\ell^{(p)\top}.
\]
The additional terms are the donor--recipient and recipient--recipient cross-products. Scores are summed separately in each imputation before averaging, giving \(\bar{\widetilde U}_j=m^{-1}\sum_p\widetilde U_j^{(p)}\), since donor assignments can differ across imputations.

\paragraph{Conditional covariance.}
Suppress the imputation index, fix the analysis and imputation parameters, and suppose donor assignments are made without using the donors' residuals. Let \(\mathcal F\) contain the predictors, observation indicators, and donor assignments, and write \(\mu_j=E(\widetilde U_j\mid\mathcal F)\). If distinct donor groups are conditionally independent, then
\begin{equation}
\operatorname{Var}\left(\sum_i U_i\mid\mathcal F\right)
=\sum_j\operatorname{Var}(\widetilde U_j\mid\mathcal F)
=\sum_j\{E(\widetilde U_j\widetilde U_j^\top\mid\mathcal F)-\mu_j\mu_j^\top\}.
\label{pmm-source-covariance}
\end{equation}
When \(\mu_j=0\), the outer products \(\widetilde U_j\widetilde U_j^\top\) have the required conditional expectation. Convergence of their normalized sum also requires a law of large numbers, with bounds on moments and donor reuse so that no donor dominates. If the conditional means are nonzero, the second moments contain \(\sum_j\mu_j\mu_j^\top\), whereas unconditional variance includes \(\operatorname{Var}(\sum_j\mu_j)\). The difference between these terms is one source of approximation error when using the summed scores without subtracting their conditional means.

\paragraph{Linear response analysis.}
In settings S1 and S2, write \(Z_j=X_j+\epsilon_j\), with \(E(\epsilon_j\mid X_j,A_j,j\in\mathcal O_Z)=0\). Let \(c_i\) be the analysis inclusion indicator, equal to one for all records in S2, and set \(J_i=i\) for observed records. At analysis slope one, omitting the common residual-variance scale,
\[
U_i=c_iX_i(Z_{J_i}-X_i)
=c_iX_i\epsilon_{J_i}+c_iX_i(X_{J_i}-X_i).
\]
It follows exactly that
\[
\widetilde U_j=w_j\epsilon_j+B_j,\qquad
w_j=\sum_{i\in\mathcal I_j}c_iX_i,\qquad
B_j=\sum_{i:J_i=j,\,i\ne j}c_iX_i(X_j-X_i).
\]
With known matching on \(X\) and observation independent of the residual conditional on the predictors, the conditional variance of the sum \(\sum_jw_j\epsilon_j\) is \(\sum_jw_j^2\operatorname{Var}(\epsilon_j\mid\mathcal F)\). This expression allows heteroscedasticity and retains the recipient-specific weights. The same donor residual appears in every recipient score, as in nearest-neighbor variance calculations \citep{yang2020pmm}.

The second moments of \(\widetilde U_j\) and \(w_j\epsilon_j\) are close when matching discrepancies are small. If \(|c_iX_i|\leq C\), donor \(j\) has \(K_j\) recipients, and \(d_n=\max_{i\in\mathcal M_Z}|X_{J_i}-X_i|\), then
\[
\frac{1}{n}\sum_jB_j^2
\leq C^2d_n^2\frac{1}{n}\sum_jK_j^2=o_p(1)
\]
when \(d_n=o_p(1)\) and \(n^{-1}\sum_jK_j^2=O_p(1)\). If \(n^{-1}\sum_jw_j^2\epsilon_j^2=O_p(1)\), Cauchy--Schwarz then gives
\[
\left|\frac{1}{n}\sum_j\{\widetilde U_j^2-w_j^2\epsilon_j^2\}\right|
\leq 2\left(\frac{1}{n}\sum_jw_j^2\epsilon_j^2\right)^{1/2}
\left(\frac{1}{n}\sum_jB_j^2\right)^{1/2}
+\frac{1}{n}\sum_jB_j^2=o_p(1).
\]
Neglecting the discrepancy in the root-\(n\) score sum requires the further condition \(n^{-1/2}\sum_jB_j=o_p(1)\); for bounded covariates, \(d_n=O_p(\log n/n)\) is sufficient. Under these conditions, the scores summed by donor give the same limiting second moment as the weighted donor residuals.

\paragraph{Logistic analysis with an imputed inclusion criterion.}
For the GS generating model, retain \(g_i(a)\) defined above. Within a donor group, decompose the score as
\[
U_i=g_i(A_j)+e_i,\qquad
e_i=\mathbf 1\{A_j>c_0\}z(A_j)\{D_i-\rho_i(A_j)\}.
\]
Conditional on \(A_j\) and \(\mathcal F\), the residuals \(D_i-\rho_i(A_j)\) are independent and have mean zero; the observed donor uses its own \(\rho_j\). Set
\[
Q_j(a)=\sum_{i\in\mathcal I_j}g_i(a),\qquad
V_i(a)=\mathbf 1\{a>c_0\}z(a)z(a)^\top\rho_i(a)\{1-\rho_i(a)\}.
\]
The law of total variance gives
\[
\operatorname{Var}(\widetilde U_j\mid\mathcal F)
=\operatorname{Var}\{Q_j(A_j)\mid\mathcal F\}
+E\left\{\sum_{i\in\mathcal I_j}V_i(A_j)\mid\mathcal F\right\}.
\]
In particular, for distinct records sharing donor \(j\),
\[
\operatorname{Cov}(U_i,U_\ell\mid\mathcal F)
=\operatorname{Cov}\{g_i(A_j),g_\ell(A_j)\mid\mathcal F\}.
\]
Thus sharing a donor value induces covariance even when the binary outcomes are conditionally independent. The outer product of the summed scores contains both the cross-products from the shared donor value and the individual binary-outcome terms. As \eqref{pmm-source-covariance} shows, subtracting \(\mu_j\mu_j^\top\) is required for the conditional covariance when \(\mu_j\ne0\); a score with marginal mean zero can have a nonzero conditional mean.

\paragraph{Combining the variance components.}
Using \(\widehat\kappa\) and \(d_i^{(p)}\) from Section~\ref{extension-2}, define
\[
\widehat\eta_i=\bar{\widetilde U}_i+\widehat\kappa\bar d_i,
\qquad
\widehat W_d=\frac{1}{nm}\sum_{i,p}
(d_i^{(p)}-\bar d_i)(d_i^{(p)}-\bar d_i)^\top.
\]
Because \(m^{-1}\sum_p(d_i^{(p)}-\bar d_i)=0\), expansion around \(\bar d_i\) shows that
\[
\widehat\Lambda
=\frac{1}{nm}\sum_{i,p}d_i^{(p)}d_i^{(p)\top}
=P_n(\bar d_i\bar d_i^\top)+\widehat W_d.
\]
Expanding the other outer product gives
\begin{align*}
P_n(\widehat\eta_i\widehat\eta_i^\top)
&=\widehat\Omega_{\mathrm{PMM}}
+\widehat\kappa P_n(\bar d_i\bar d_i^\top)\widehat\kappa^\top\\
&\quad+\widehat\kappa P_n(\bar d_i\bar{\widetilde U}_i^\top)
+P_n(\bar{\widetilde U}_i\bar d_i^\top)\widehat\kappa^\top.
\end{align*}
The last two terms equal \(\widehat C_{\mathrm{PMM}}\). Substituting the expression for \(\widehat\Lambda\) therefore proves the exact decomposition
\begin{align*}
\widehat\Delta_{\mathrm{PMM}}
&=\widehat\Omega_{\mathrm{PMM}}
+\widehat\kappa\widehat\Lambda\widehat\kappa^\top
+\widehat C_{\mathrm{PMM}}\\
&=P_n(\widehat\eta_i\widehat\eta_i^\top)
+\widehat\kappa\widehat W_d\widehat\kappa^\top.
\end{align*}
Both matrices on the last line are positive semidefinite. Thus, when \(\widehat\tau\) is invertible, the resulting covariance is also positive semidefinite:
\[
\widehat\Gamma_{\mathrm{PMM}}
=\frac{1}{n}\widehat\tau^{-1}
\widehat\Delta_{\mathrm{PMM}}(\widehat\tau^{-1})^\top.
\]
The decomposition also explains why averaging \(d_i^{(p)}d_i^{(p)\top}\) is not identical to taking the outer product of \(\bar d_i\).

These identities establish the algebra of the proposed variance estimator. The preceding calculations give its cross-moment limit and the covariance induced by donor reuse under the stated model assumptions. A consistency proof for the full fitted PMM estimator would additionally need to account for estimated matching coefficients, conditional score means, and imputation randomization. We evaluate the approximation through estimated standard errors and confidence-interval coverage in the simulation settings studied.

\section{Additional Simulation Results}
\label{appendix-simulation-figures}
Tables~\ref{tab:pmm-k-sensitivity-supp} and \ref{tab:pmm-k4000-supp} give PMM results for donor pool sizes \(k=20,30\). Each cell has 2500 replications. Bias and coverage of \(\beta_0\) use \(\beta_0=0.885377976781648\); coverage of \(\beta^\ast\) is approximated using the mean estimate within each cell. RW/RR pairs refer to RW-PMM/RR-PMM.

\begin{table}[ht]
\centering
\scriptsize
\setlength{\tabcolsep}{3.5pt}
\begin{tabular}{llrrrrrr}
\toprule
$m$ & \makecell{Imputation\\model} & \makecell{Mean\\estimate} & Bias &
\makecell{Empirical\\SD} & \makecell{Mean estimated SE\\RW/RR} &
\makecell{Coverage of $\beta^\ast$\\RW/RR} & \makecell{Coverage of $\beta_0$\\RW/RR}\\
\midrule
5 & PMM ($k=20$) & 0.838 & -0.048 & 0.137 & 0.131 / 0.172 & 0.940 / 0.978 & 0.904 / 0.965 \\
5 & PMM ($k=30$) & 0.823 & -0.062 & 0.135 & 0.130 / 0.172 & 0.938 / 0.982 & 0.902 / 0.960 \\
\addlinespace[2pt]
25 & PMM ($k=20$) & 0.837 & -0.049 & 0.124 & 0.124 / 0.167 & 0.946 / 0.989 & 0.924 / 0.979 \\
25 & PMM ($k=30$) & 0.820 & -0.065 & 0.132 & 0.124 / 0.167 & 0.932 / 0.986 & 0.884 / 0.961 \\
\addlinespace[2pt]
50 & PMM ($k=20$) & 0.834 & -0.051 & 0.127 & 0.124 / 0.167 & 0.938 / 0.987 & 0.919 / 0.978 \\
50 & PMM ($k=30$) & 0.817 & -0.068 & 0.127 & 0.123 / 0.167 & 0.939 / 0.989 & 0.900 / 0.968 \\
\addlinespace[2pt]
100 & PMM ($k=20$) & 0.833 & -0.052 & 0.127 & 0.123 / 0.166 & 0.949 / 0.991 & 0.913 / 0.979 \\
100 & PMM ($k=30$) & 0.821 & -0.064 & 0.127 & 0.124 / 0.167 & 0.945 / 0.990 & 0.896 / 0.972 \\
\bottomrule
\end{tabular}
\caption{GS PMM results at \(n=2000\), with 500 validated records, for \(k=20,30\).}
\label{tab:pmm-k-sensitivity-supp}
\end{table}

\begin{table}[ht]
\centering
\scriptsize
\setlength{\tabcolsep}{3.5pt}
\begin{tabular}{llrrrrrr}
\toprule
$m$ & \makecell{Imputation\\model} & \makecell{Mean\\estimate} & Bias &
\makecell{Empirical\\SD} & \makecell{Mean estimated SE\\RW/RR} &
\makecell{Coverage of $\beta^\ast$\\RW/RR} & \makecell{Coverage of $\beta_0$\\RW/RR}\\
\midrule
5 & PMM ($k=20$) & 0.866 & -0.020 & 0.096 & 0.095 / 0.120 & 0.938 / 0.973 & 0.934 / 0.968 \\
5 & PMM ($k=30$) & 0.851 & -0.034 & 0.095 & 0.094 / 0.119 & 0.943 / 0.977 & 0.920 / 0.960 \\
\addlinespace[2pt]
25 & PMM ($k=20$) & 0.864 & -0.022 & 0.091 & 0.090 / 0.116 & 0.948 / 0.984 & 0.941 / 0.978 \\
25 & PMM ($k=30$) & 0.853 & -0.032 & 0.092 & 0.090 / 0.117 & 0.942 / 0.988 & 0.924 / 0.979 \\
\addlinespace[2pt]
50 & PMM ($k=20$) & 0.866 & -0.020 & 0.091 & 0.090 / 0.116 & 0.947 / 0.986 & 0.933 / 0.983 \\
50 & PMM ($k=30$) & 0.853 & -0.032 & 0.091 & 0.089 / 0.116 & 0.945 / 0.989 & 0.924 / 0.980 \\
\addlinespace[2pt]
100 & PMM ($k=20$) & 0.865 & -0.020 & 0.090 & 0.089 / 0.116 & 0.946 / 0.986 & 0.937 / 0.982 \\
100 & PMM ($k=30$) & 0.857 & -0.028 & 0.091 & 0.089 / 0.116 & 0.944 / 0.987 & 0.924 / 0.979 \\
\bottomrule
\end{tabular}
\caption{GS PMM results at \(n=4000\), with 1000 validated records, for \(k=20,30\). Columns are defined as in Table~\ref{tab:pmm-k-sensitivity-supp}.}
\label{tab:pmm-k4000-supp}
\end{table}
\FloatBarrier

\begin{figure}[ht]
\centering
\includegraphics[width=\textwidth]{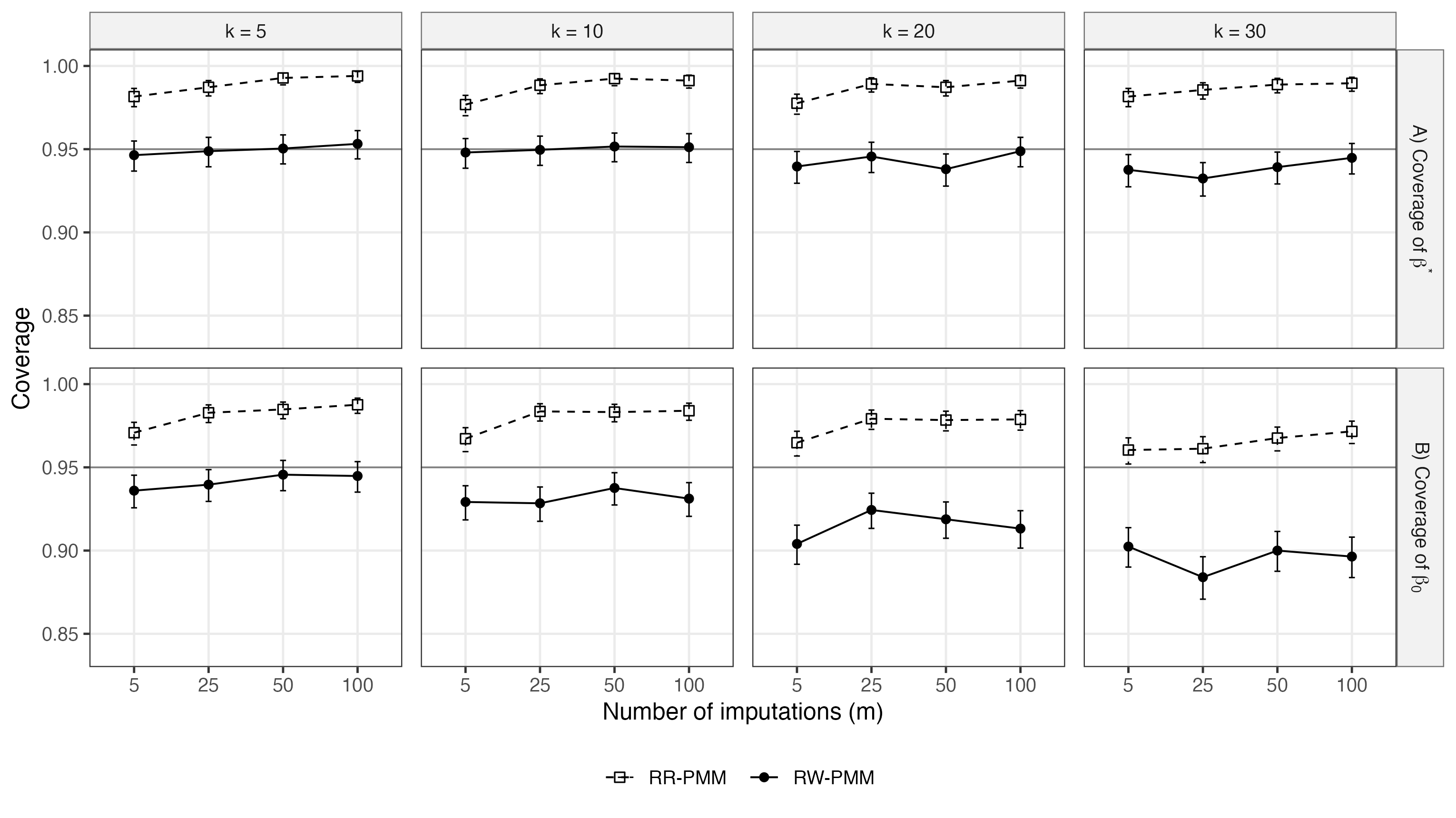}
\caption{GS PMM coverage at \(n=2000\), with 500 validated records and 2500 replications per cell. Columns show donor pool size \(k\); rows show coverage of \(\beta^\ast\) and coverage of \(\beta_0\).}
\label{fig:gs-pmm-current-n2000-supp}
\end{figure}

\begin{figure}[ht]
\centering
\includegraphics[width=\textwidth]{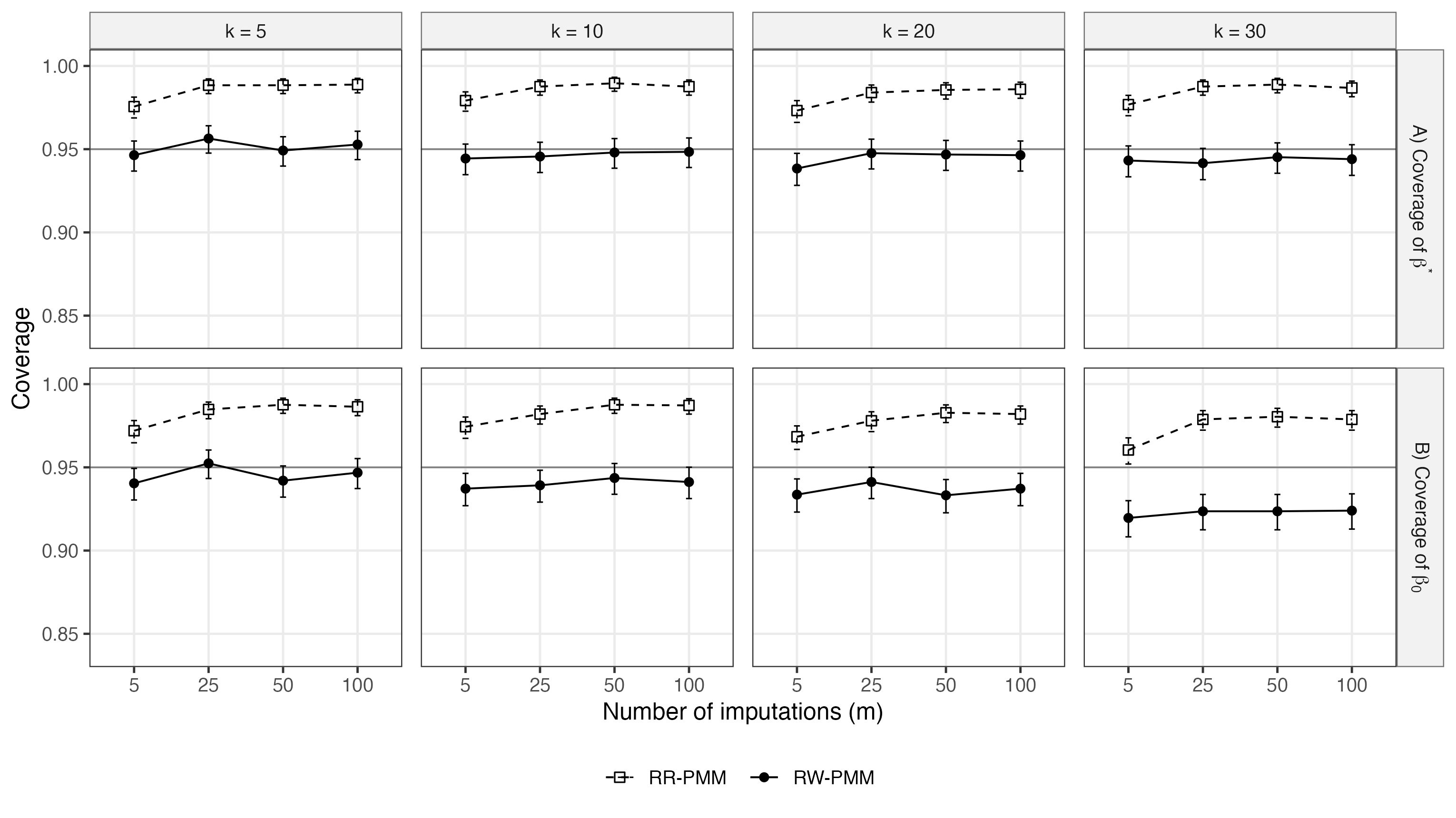}
\caption{GS PMM coverage at \(n=4000\), with 1000 validated records and 2500 replications per cell. Columns show donor pool size \(k\); rows show coverage of \(\beta^\ast\) and coverage of \(\beta_0\).}
\label{fig:gs-pmm-current-n4000-supp}
\end{figure}

Figures~\ref{fig:gs-pmm-current-n2000-supp} and \ref{fig:gs-pmm-current-n4000-supp} display both coverage measures across donor pool sizes. The horizontal line marks 0.95. Error bars are nominal 95\% binomial Monte Carlo intervals; those for coverage of \(\beta^\ast\) are approximate because they use the mean point estimate from the same replications.

\begin{figure}[ht]
\centering
\includegraphics[width=\textwidth]{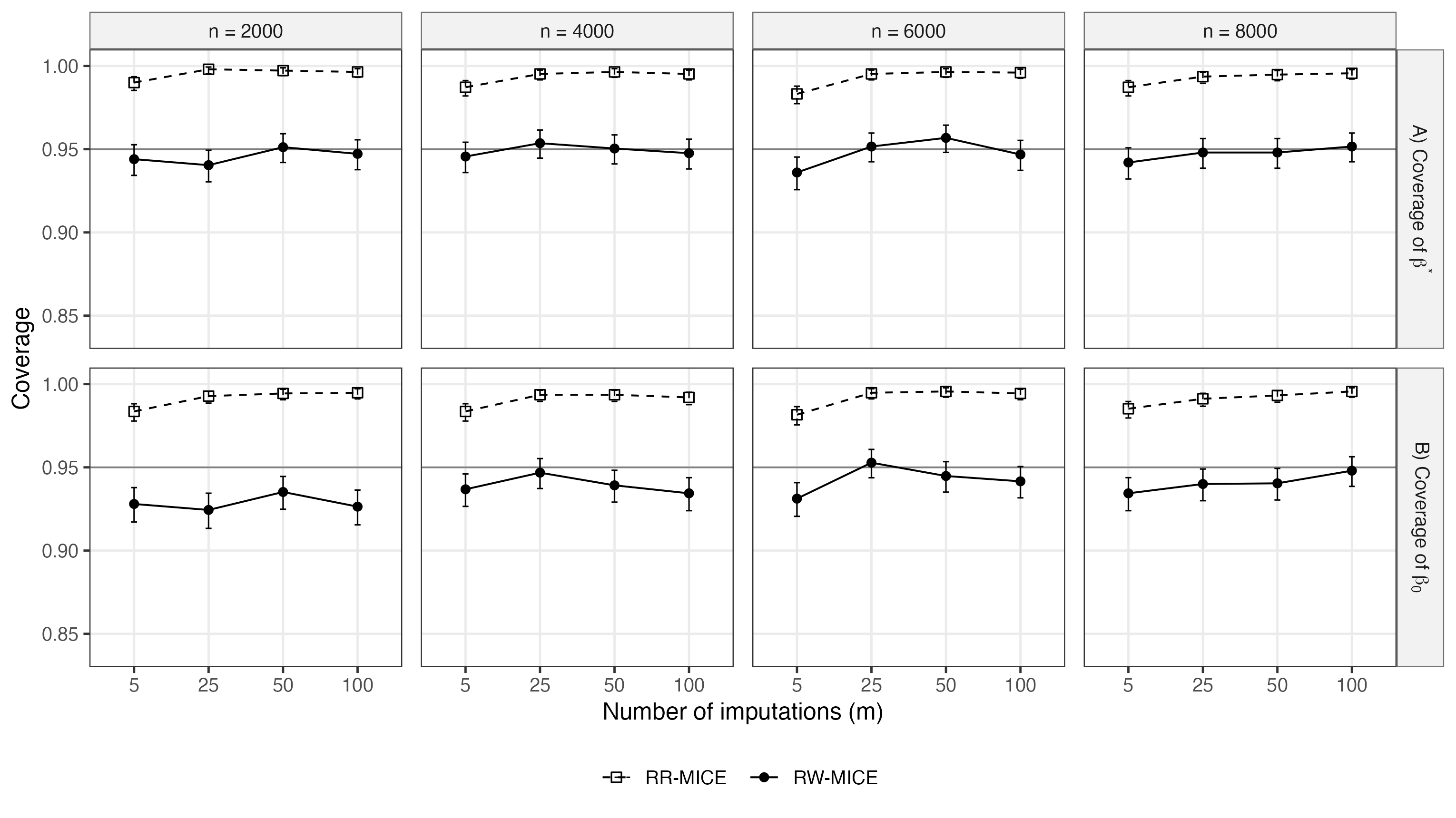}
\caption{GS parametric MICE coverage across sample sizes, with 2500 replications per cell. Rows show coverage of \(\beta^\ast\) and coverage of \(\beta_0\). RW and RR use the same imputations. Error bars are defined above.}
\label{fig:gs-parametric-full-supp}
\end{figure}

Figure~\ref{fig:gs-parametric-full-supp} shows the parametric MICE results for \(n=2000,4000,6000,8000\), with \(n/4\) validated records at each sample size.
\clearpage

\bibliography{papers}

\end{document}